\documentclass[11pt,journal,draftcls,onecolumn,a4paper]{IEEEtran}

\pdfoutput=1
\def\articleofficiel{1}
\ifx\articleofficiel\undefined
\fi

\usepackage{ifthen}
\usepackage{graphicx}%
\usepackage{multirow}
\usepackage{amsmath}%
\usepackage{amssymb}%
\usepackage{amsfonts}%
\usepackage{bbm}
\usepackage{textcomp}
\usepackage{color}
\usepackage[latin1]{inputenc}
\usepackage{enumerate}
\usepackage{subfigure}
\newcounter{assumption}
\newenvironment{assumption}
  {\renewcommand{\theenumi}{(H-\arabic{assumption})}%
  \setlength{\leftmargini}{4em}\refstepcounter{assumption}
  \begin{enumerate}}
  {\end{enumerate}}

\def\rme{\mathrm{e}}    % Fonction exponentielle
\def\rmi{\mathrm{i}}    % Imaginaire Pur
\def\rset{\mathbb{R}} % Réels
\def\PP{\mathbb{P}}             % Espaces de proba
\def\E{\mathbb{E}}              % Espérance d'une var
\def\var{\mathrm{Var}}
\def\1{\mathbbm{1}}
\DeclareMathOperator*{\argmin}{Arg\,min}

\newcommand{\iid}{independent and identically distributed}

\newtheorem{proposition}{Proposition}[section]

\newtheorem{theorem}{Theorem}[section]

\newtheorem{remark}{Remark}[section]
{\textbf{#1} }%
{\begin{flushright}$\blacksquare$ \end{flushright}}
\def\rme{\mathrm{e}}    % Fonction exponentielle
\def\rmi{\mathrm{i}}    % Imaginaire Pur
\def\rset{\mathbb{R}} % Réels
\def\PP{\mathbb{P}}             % Espaces de proba
\def\E{\mathbb{E}}              % Espérance d'une var

\usepackage{ifthen}
\newcommand{\eqsp}{\;}

\newcommand{\eqdef}{\ensuremath{\stackrel{\mathrm{\Delta}}{=}}}

\begin{document}

\title{Semiparametric curve alignment and shift density estimation for biological data}

\author{T. Trigano, U. Isserles and Y. Ritov}

\markboth{IEEE Transactions in Signal Processing}{Trigano \MakeLowercase{\textit{et al.}}: Semiparametric curve alignment and shift density estimation with application to biological data}

\maketitle

\begin{abstract}
Assume that we observe a large number of signals, all of them with identical, although unknown, shape, but with a different random shift. The objective is to estimate the individual time shifts and their distribution. Such an objective appears in several biological applications like neuroscience or ECG signal processing, in which the estimation of the distribution of the elapsed time between repetitive pulses with a possibly low signal-noise ratio, and without a knowledge of the pulse shape is of interest.  We suggest an M-estimator leading to a three-stage algorithm: we first split our data set in blocks, then the shift estimation in each block is done by minimizing a cost function based on the periodogram; the estimated shifts are eventually plugged into a standard density estimator. We show that under mild regularity assumptions the density estimate converges weakly to the true shift distribution. The theory is applied both to simulations and to alignment of real ECG signals. The proposed approach outperforms the standard methods for curve alignment and shift density estimation, even in the case of low signal-to-noise ratio, and is robust to numerous perturbations common in ECG signals.
\end{abstract}

\section{Introduction}

We investigate in this paper a specific class of stochastic nonlinear inverse problems. We observe a collection of \(M+1\) uniformly sampled signals in a finite interval $[0,T]$
\begin{equation}
y_{j}(t_i) = s(t_i-\theta_j) + \sigma \varepsilon_j(t_i),\ t_i\in [0,T], \ j=0\ldots M  \label{eq:shifted_model_intro}
\end{equation}
where $s$ is an unknown signal, $\{\theta_j, j = 0 \ldots M\}$ are independent real-valued continuous random variables with common probability density function $f$ which represent a shift parameter, and $\varepsilon_0,\dots,\varepsilon_M$ are independent standard white noise processes with standard deviation \(\sigma\) and independent of $\theta_0,\dots,\theta_M$. Our aim is to estimate either $\{\theta_j, j = 0 \ldots M\}$, or the shift distribution $f$. Similar models appear commonly in practice in numerous fields. For instance, a common problem in functional data analysis (FDA) is to align curves obtained in a series of experiments with varying time shifts, before extracting their common features; we refer to~\cite{silverman:ramsay:2005} and~\cite{ferraty:vieu:2006} for an in-depth discussion on the problem of curve alignment in FDA applications. In data mining applications, after splitting the data into different homogeneous clusters, observations of a same cluster may differ. Such variations take into account the variability of individual waveforms inside one given group. In the framework described by \eqref{eq:shifted_model_intro}, the knowledge of the translation parameter $\theta$, and more specifically of its distribution, can be used to determine the inner variability of a given cluster of curves. Several papers (e.g.~\cite{ramsay:1998,ramsay:li:1998,ronn:2001,gasser:kneip:1995,kneip:gasser:1992}) focus on this specific model for many different applications in biology or signal processing. 
%
%The problem we have to tackle can be seen as an inverse problem. Several authors have investigated nonparametric maximum likelihood estimation for stochastic inverse problems, using variants of the Expectation Maximization (EM) algorithm such as~\cite{chafai:loubes:2006}. In our framework, the function $s$ is unknown, thus forbidding the use of such techniques. 
Such a problem can also be related to curve alignment problems as in~\cite{silverman:ramsay:2005}, which are typically encountered in medicine (growth curves) and traffic data. Many methods previously introduced rely on a preliminary estimation of $s$, thus introducing an additional error in the estimation of $\{\theta_j, j = 0 \ldots M\}$. For example,~\cite{gasser:kneip:1995} proposed to estimate the shifts by aligning the maxima of the curves, their position being approximated by the zeros of a kernel estimate of the derivative. Similar discussions can  be found in recent contributions in the system identification framework, e.g.~\cite{pawlak:2007,greblicki:2008,hasiewicz:2009}. In particular,~\cite{greblicki:2008} provides a two-stage algorithm which estimates jointly a parametric component and a functional. Since we do not rely on any information or estimation regarding $s$ in this paper, it is of interest to consider it as a nuisance parameter, and the shifts $\{\theta_j, j = 0 \ldots M\}$ (or $f$) as a parameter of interest, and consider~\eqref{eq:shifted_model_intro} as a semiparametric model as described in~\cite{bickel:ritov:1998}. However, if one is interested in $s$ while having estimates of the shifts $\hat\theta_1,\dots,\hat\theta_M$, one can easily proceed and use $\hat s(t) = M^{-1}\sum_{j=0}^M y_j(t+\hat\theta_j)  $ as an estimate of the signal $s$.

Our contribution is close to recent shift estimation techniques described in~\cite{vimond:2010} and~\cite{castillo:2009}. Both rely on the fact that the spectral density of one given signal remains invariant by shifting, and therefore, it is well fitted for semiparametric methods when $s$ is unknown. In~\cite{vimond:2010}, the problem addressed is the joint estimation of $K$ shifts parameters, when $K$ is a fixed number of curves (unlike what is done in the current paper where $K\to\infty$). This leads to a semiparametric estimation technique similar to the papers of~\cite{gamboa:loubes:maza:2005,lavielle:levyleduc:2005}. The advantage of such an estimator is that it is asymptotically efficient, consistent and asymptotically normal. However, when the number of curves to process is important, the method leads to a computationally intensive optimization problem. It is therefore of interest in practical applications to deal with blocks of smaller size which include one identical reference curve, as done in this paper. In~\cite{castillo:2009}, the authors estimate the shift probability density function when the number of curves is infinite, but the corresponding alignment procedure is performed one curve after the other, by means of the minimization of a penalized likelihood function. Such an approach makes sense when we have a few curves to compare, but when we dispose of many signals, the shifts parameters may be estimated jointly and more efficiently. % The asymptotics for an increasing number of curves is presented in this paper.

% The power spectral density of one given curve remains invariant under shifting, and therefore, it is well fitted for semiparametric methods when $s$ is unknown or the variance of the noise is high. Methods described in~\cite{gamboa:loubes:maza:2005} or in~\cite{lavielle:levyleduc:2005} are based on filtered power spectrum information, and are relevant if the number of curves to reshift is small, which is the case in some applications, such as traffic forecasting. The authors show that their estimator is consistent and asymptotically normal, however, this asymptotic study is done when the number of samples for each curve tends to infinity, the number of curves remaining constant and usually small. On the other hand, it is of interest to investigate the asymptotics for an increasing number of curves, since the duration of the experiment can be more easily controlled than the amount of noise per unit. The asymptotics for an increasing number of curves is presented in this paper.

% \begin{figure}
% \centering
% \includegraphics[width=0.8\linewidth]{ECG_makori}
% \caption{Example of ECG signal from the MIT-BIH database.}\label{fig:example_signal}
% \end{figure}

In our main application, we analyze ECG signals. % In recordings of the heart's electrical activity, at each cycle of contraction and release of the heart muscle, we get a characteristic P-wave, which depicts the depolarization of the atria, followed by a QRS complex stemming from the depolarization of the ventricles and a T-wave corresponding to the repolarization of the heart muscle. We refer to \cite[Chapter 12]{guyton:hall:1996} for an in-depth description of the heart cycle. A typical ECG signal is shown in Figure~\ref{fig:example_signal}. Different positions of the electrodes, transient conditions of the heart, as well as some malfunctions and several perturbations (baseline wander, powerline interference), can alter the shape of the signal. 
We aim at situations where the heart electrical activity remains regular enough in the sense that the shape of each cycle remains approximately repetitive, so that after prior segmentation of the ECG recording, the above model still holds. This is the case for heart malfunctions such as sinus or supraventricular tachycardia, as mentioned in~\cite{guyton:hall:1996}. This preliminary segmentation can be done efficiently, for example, by taking segments around the easily identified maxima of the QRS complex, as it can be found in~\cite{gasser:kneip:1995}, or by means of digital filters as suggested in~\cite{pan:tomkins:1985}. It is of interest to estimate $\{\theta_j, j = 0 \ldots M\}$, in~\eqref{eq:shifted_model_intro}, since these estimates can be used afterwards for a more accurate estimation of the heart rate distribution.
% In normal cases, such estimation can be done accurately by using some common FDA methods (e.g. using only the initial segmentations). However, when the activity of the heart is more irregular, a more precise alignment can be helpful. This happens for example in cases of cardiac arrythmias, whose identification can be easier if the heart cycles are accurately aligned.
 Another measurement often used by cardiologists is the mean ECG signal. A problem encountered in that case is that improperly aligned curves can yield an average on which the characteristics of the ECG cycle are lost. The proposed method leads to a more efficient estimation of the mean cycle by averaging the segments after an alignment according to a well-estimated shift.

The paper is organized as follows. Section~\ref{sec:estimator} describes the assumptions made and the method to derive the estimators of the shifts and of their distribution. This method is based on the optimization of a cost function, based on the comparison between the power spectrum of the average of blocks of curves and the average of the individual power spectra. Since we consider a large number of curves, we expect that taking the average signal will allow to minimize the cost criterion consistently. We provide in Section~\ref{sec:convergence} theoretical results on the efficiency of the method and on the weak convergence of the density estimate. In Section~\ref{sec:applications}, we present simulations results, which show that the proposed algorithm performs well for density estimation, and study its performances under different conditions. We also applied the methodology to the alignment of ECG signals, and show that the proposed algorithm outperforms the standard FDA methods. Proofs of the discussed results are presented in the appendix.

 % 4.3
\section{Nonparametric estimation of the shift distribution}
\label{sec:estimator}

In this section, we state the main assumptions that will be used in the rest of the paper, and propose an algorithm which leads to an M-estimator of the shifts. Using these estimators, we obtain a plug-in estimate of the shift probability density function.

\subsection{Assumptions}

%We consider a discrete version of \eqref{eq:shifted_model_intro}. That is, 
Assume that we observe $M+1$ sampled noisy signals on a finite time interval $[0,T]$, each one being shifted randomly by $\theta$; a typical signal is expressed by
\begin{equation}\begin{split}
&y_{j}(t_i) = s(t_i-\theta_j) + \sigma \varepsilon_j(t_i),\
\\
&t_i=\frac{(i-1)T}{n}, \quad  i=1\ldots n, \ j=0\ldots M , \label{eq:shifted_model}
\end{split}\end{equation}
where the processes $\{\varepsilon_j,j=0\ldots M\}$ are assumed to be standard Gaussian white noises, and the  variance $\sigma^2$ is assumed to be constant. We also assume that the whole signal is within the sampling frame, which can be formalized by the following assumption:
\begin{assumption}
\item The distribution of $\theta$ and the shape $s$ both have bounded non-trivial support,    $[0,T_\theta]$ and $[0,T_s]$, respectively, \label{hyp:finite_support}
and $T_\theta + T_s < T$. \label{hyp:unrecovering}
\end{assumption}
As pointed out in~\cite{ritov:1989}, under this assumption we can consider $s$ as a periodic function with associated period $T$. Without any loss of generality, we further assume that $T\eqdef 2\pi$ in order to simplify notations. We also assume:
\begin{assumption}
\item \label{hyp:independence}
$s\in L^2 ([0,T_s])$ and its derivative \(s'\in L^{\infty}\). \label{hyp:finite_energy}
\end{assumption}
Assumption~\ref{hyp:finite_support} implies that we observe a sequence of identical curves with additive noise, so that the spectral information is the same for all curves. Assumption~\ref{hyp:finite_energy} is critical to guarantee the existence of the Energy Spectral Density (ESD) of the studied curve and of the terms appearing in later sections. Note that the boundedness of the derivative is assumed for the sake of convenience (in order to show easily that the discretization error in the later parts can be neglected); the proposed method would also give good results on curves showing discontinuities. We finally make the following assumptions on the random variables appearing in~(\ref{eq:shifted_model}):
\begin{assumption}
\item
The shifts $\{\theta_j,j=0\ldots M\}$ are continuous random variables, independent and identically distributed with common probability density function $f$ which is assumed to be uniformly bounded. We also consider the first shift $\theta_0$ as known, and without loss of generality we fix $\theta_0 \eqdef 0$. Finally, we assume that the variables $\{\varepsilon_j(t_i) , j=0,\dots,M, i=1,\dots,n \}$  are standard normal independent random variables, which are also independent of $\{\theta_j,j=0\ldots M\}$.
\label{hyp:ind}
\end{assumption}

%and align all the curves with respect to $y_0$.
% Assumption \ref{hyp:ind} is stronger than needed and it is given for the sake of simplicity. We need only to assume  that the processes  $\{\varepsilon_j(t_1),\dots,\varepsilon_j(t_n)\}$  are independent and mixing. See the discussion following \eqref{eq:discrete_model_fourier} for more details.

\subsection{Computation of the shift estimators and of their density}

The intuitive idea of the proposed algorithm is as follows. Assume, for the sake of the argument, that $\sigma=0$; then, when the shifts are known and corrected, the individual signals are equal to their average. Consequently, the average of their ESDs is equal to the ESD of the mean signal. On the other hand, if the shifts are not corrected, then the average signal is a convolution of the original shape with the shift distribution, and hence its ESD is strictly different from the average of the individual ESD's. 

\begin{figure}[at]
\centering
\scalebox{0.55}{\input{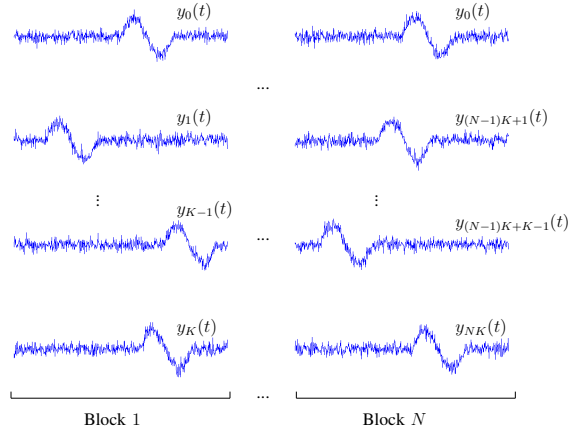}}
\caption{Split of the curves data set}\label{fig:data-splitting}
\end{figure}

Following the method of~\cite{castillo:2009}, we propose to plug estimators of $\{ \theta_j$, $j=1\ldots M \}$ into an estimate of $f$. We start by splitting our dataset in $N$ blocks of $K+1$ curves each, as shown in Figure~\ref{fig:data-splitting}. Observe that $y_0$ is included in each block, since all the rest of the signals are aligned with it. The motivation to split the dataset into smaller blocks is twofold: it reduces the variance of the estimators of the shifts by estimating them jointly, and also provides smooth functions for the optimization procedure detailed in this section. The first step is therefore to estimate the vectors of shifts $\{\boldsymbol\theta_m, m=1\ldots N\}$, where for all integer $m$, $\boldsymbol{\theta}_m \eqdef (\theta_{(m-1)K+1},\ldots,\theta_{m\,K}) $.

The estimation of $\boldsymbol{\theta}_m$ is achieved by minimizing a cost function. For any continuous-time and $2\pi$-periodic signal $y$, we denote by $S_y$ its energy spectral density, that is for all $\omega$:
\begin{equation}
\label{psdr}
S_y(\omega) \eqdef  \left| \frac{1}{2\pi} \int_{0}^{2\pi} y(t) \rme^{-i\omega t}\, dt \right|^2\eqsp .
\end{equation}
This quantity is of interest, since it remains invariant by shifting. For each integer $m=1\ldots N$, we define the mean of $K$ signals translated by some correction terms $\boldsymbol{\alpha}_m \eqdef (\alpha_{(m-1)K+1},\ldots,\alpha_{mK})$:
\begin{align}\label{eq:reshifted}
&\bar{y}_m (t ; \boldsymbol{\alpha}_m)
\\
&\eqdef
\frac{1}{K+\lambda}\left( \lambda
  y_0(t) + \sum_{l=(m-1)K+1}^{mK} y_l (t + \alpha_l) \right)
,\nonumber
\end{align}
% \begin{equation}
% \bar{y}_n \eqdef \frac{1}{K+\lambda(K)}\left( \lambda(K) y_0 + \sum_{k=(n-1)K+1}^{nK} y_k \right)\label{eq:mean_signal}
% \end{equation}
% the weighted average process taken for the $K+1$ curves of block
% $n$,
where $\lambda\eqdef\lambda(K)$ is a positive number which depends on $K$, and is introduced in order to give more importance to the reference signal $y_0$. For any \(m=1,\dots,N\) we now consider:
\begin{equation}
\frac{1}{M+1} \sum_{l=0}^{M} S_{y_{l}} (\omega) - S_{\bar{y}_{m} (\cdot; \boldsymbol{\alpha}_m)} (\omega) \eqsp . \label{eq:quantity}
\end{equation}
The function described in~(\ref{eq:quantity}) represents the difference between the mean of the ESDs and the ESD of the average signal of the \(m\)-th block. Since the observed signals are sampled, the integral of $S_y (\omega)$ will be in practice approximated by its Riemann sum, that is
$$
\hat{S}_y (k) = \left| \frac{1}{n}  \sum_{m=1}^{n} y(t_m) \rme^{-
  {2\rmi \pi m k}/{n}} \right|^2,\quad k\in{\cal K} \eqsp ,
  $$ 
where ${\cal K} = \{-\frac{n-1}2, -\frac{n-3}2,\dots,\frac{n-1}2\}$ (note that $k$ in the latter is not necessarily an integer). Let the sequence \(C_m(\boldsymbol{\alpha}_m) \eqdef \{C_m(k,\boldsymbol{\alpha}_m):k\in{\cal K}\}\) be defined by
\begin{equation}
C_m(k,\boldsymbol{\alpha}_m ) \eqdef \frac{1}{M+1} \sum_{l=0}^{M} \hat{S}_{y_{l}}(k) - \hat{S}_{\bar{y}_m (\cdot ; \boldsymbol{\alpha}_m) } (k) \eqsp , \label{eq:aux_function}
\end{equation}
and let $\{\nu_k, k \in \cal K \}$ be a sequence of nonnegative numbers such that \(\nu_{-k}=\nu_k\) and \(\sum_k k^2 \nu_k<\infty\) when $n$ tends to infinity. The proposed M-estimator of $\boldsymbol{\theta}_m$ is denoted by $\boldsymbol{\hat{\theta}}_{m}$ and is given by
\begin{equation}
{\hat{\boldsymbol\theta}_{m}} \eqdef
\argmin_{\boldsymbol{\alpha}_m\in [0;2\pi]^K} \|C_m(\boldsymbol{\alpha}_m)\|_\nu^2\label{eq:M-estimate},
\end{equation}
where \(\|C_m(\boldsymbol{\alpha}_m)\|_\nu^2=\sum_{k\in {\cal K}}\nu_k|C_m(k,\boldsymbol\alpha_m)|^2\).
\begin{remark}
As aforementioned, all the blocks have one curve $y_0$ in common. We impose this constraint in order to address the problem of identifiability. Without this precaution, replacing ${\boldsymbol\alpha}_{m}$ by $ {{\boldsymbol\alpha}_{m}}  + (c,c,\ldots,c), \ c \in \rset $ and the signal $s(\cdot)$ by $s(\cdot-c)$ in the $m$-th block would let \eqref{eq:aux_function} invariant. Adding the curve $y_0$ in each block as a reference allows to estimate the shifts with respect to a same common reference.
\end{remark}

The estimator of the probability density function $f$, denoted by $\hat{f}_{M,h}$, is then computed by plugging the estimated values of the shifts in a known density estimator, such as the regular kernel density estimator~\cite{dasgupta:2008}, that is for all real $x$ in $[0;2\pi]$:
\begin{equation}
\hat{f}_{M,h}(x) = \frac{1}{(M+1) h} \sum_{m=0}^{M} \psi\left(
  \frac{x-\hat{\theta}_m}{h} \right)\eqsp, \label{eq:kernel_estimate}
\end{equation}
where the kernel $\psi$ is a nonnegative function integrating to $1$ with a bounded derivative and $h$ the classical bandwidth parameter of the kernel. In this paper we provide a proof of weak convergence of the empirical distribution function of $\{\hat\theta_j,\ j=1\ldots M\}$ under some mild conditions. More specifically, we shall get from Theorem~\ref{prop:weak_convergence} that $\hat{f}_{M,h}(x)$ converges pointwise to $f(x)$ when both $M\to\infty$ and $n\to\infty$.
% $$
% \frac{1}{(M+1)} \sum_{m=0}^M g(\hat{\theta}_m) \longrightarrow
% \E[g(\theta)]\eqsp,
% $$
% when both $M\to\infty$ and $n\to\infty$, for any bounded continuous function $g$ on \([0,2\pi]\).
% It should be noticed, however, that $h$ is a free parameter which may exhibit a strong influence on the resulting estimate. A too small value of $h$ leads to over-fitting, whereas taking large values of $h$ leads to hide the multimodality of $f$, if any. Choosing a data-driven bandwidth selection of $h$ is thus far from trivial, and out of the scope of this paper: we refer to~\cite{berlinet:devroye:1994,jones:marron:1996} for an in-depth description of the existing procedures. The bandwidth $h$ is chosen by Silverman's ``rule-of-thumb''~\cite{silverman:1986}.
 % 4.3

\section{Theoretical aspects}
\label{sec:convergence}

We provide in this section theoretical results on the estimators described in~\eqref{eq:M-estimate} and \eqref{eq:kernel_estimate}.
%Recall that the total number of curves is $M=NK+1$, where $N$ is the number of blocks and $K+1$ is the number of curves in each block. The first curve $y_0$ is a common reference curve which is included in all blocks.
 We denote by $c_s (k)$ the discrete Fourier transform (DFT) of $s$:
\begin{equation*}
c_s(k) \eqdef \frac{1}{n} \sum_{m=1}^{n} s(t_m) \rme^{-
  {2\rmi \pi m k}/{n}} \eqsp , k \in {\cal K} \ ,
\end{equation*}
and by $f_{k,l}$ the DFT of $y_l$:
\begin{equation*}
f_{k,l} \eqdef \frac{1}{n} \sum_{m=1}^{n} y_l(t_m) \rme^{-  {2\rmi \pi m k}/{n}} \eqsp , k \in {\cal K}, \ l=0 \ldots M \eqsp .
\end{equation*}
Let \(\theta_l=\bar\theta_l+\epsilon_l\) where \(\bar\theta_l\in\{t_1,\dots,t_n\}\) is the sampling point whose value is the closest to the actual shift $\theta_l$, and $\epsilon_l$ denotes the discretization error. Observe that since the signal is unifomrly sampled on $[0,2\pi]$, we may write that \(|\epsilon_l|<\pi/n\). Using this notation, relation~(\ref{eq:shifted_model}) becomes in the Fourier domain for all $k\in{\cal K}$ and $l=0\ldots M$:
\begin{equation}
 \label{eq:discrete_model_fourier}
\begin{split}
f_{k,l} &= \frac{1}{n} \sum_{m=1}^{n} s(t_m-\theta_l) \rme^{-
 {2\rmi \pi m k}/{n}}
  \\
  &\hspace{2em} + \frac{\sigma}{\sqrt{n}} \left(V_{k,l} + \rmi
  W_{k,l} \right)
  \\
   &= \rme^{-\rmi k \bar\theta_l}\frac{1}{n} \sum_{m=1}^{n} s(t_m-\epsilon_l) \rme^{-
 {2\rmi \pi m k}/{n}}
  \\
  &\hspace{2em} + \frac{\sigma}{\sqrt{n}} \left(V_{k,l} + \rmi
  W_{k,l} \right)
  \\
  &=\rme^{-\rmi k \theta_l} c_s(k) +\mathrm O(k n^{-1})
  %\\  &\hspace{2em}
  +\frac{\sigma}{\sqrt{n}} \left(V_{k,l} + \rmi
  W_{k,l} \right)  ,
  \end{split}
\end{equation}
where the middle equality is obtained by the shift property of the DFT and the last equality stems from Taylor-Lagrange inequality due to~\ref{hyp:finite_energy}. By the white noise assumption \ref{hyp:ind}, the sequences $\left\{V_{k,l}, \  k\in{\cal K} \right\}$ and $\left\{W_{k,l}, \  k\in{\cal K}\right\}$ in (\ref{eq:discrete_model_fourier}) are \iid\ with same standard multivariate normal distribution $\mathcal{N}_n(0,I_n)$. The \(\mathrm O(k n^{-1})\) term  is a result of the sampling operation and is purely deterministic; since it is assumed that $\sum_k k^2 \nu_k <\infty$, the contribution of this deterministic error to the cost function shall be no more than $\mathrm O(n^{-1})$, and will further on be neglected since it is not going to induce shift estimation errors greater than the length of a single bin (i.e. $n^{-1}$), while it will be shown that the statistical estimation error is \(\mathrm O_\PP(n^{-1/2})\). % Since we investigate the asymptotic properties of the estimate, especially when both $n$ and $K$ tend to infinity, we hereafter consider this discretization error as negligible and ignore it.

Note that Assumption \ref{hyp:ind} might be considered too strong, and we indeed state it for simplicity. It can be weakened to include more general random variables $\varepsilon_j(t_1),\dots,\varepsilon_j(t_n)$, as long as the Central Limit Theorem can be applied in \eqref{eq:discrete_model_fourier}. The $\sqrt n$ term appearing in this equation should be then understood as the normalization constant of the mean error in the $k$-th tap. In particular the homogeneity of the noise distribution is not needed, and some dependency may be permitted as long as the sequence remains with some mixing property. The process should essentially be such that for any $0<a<b<T$, $\max_{j} \var ((b-a)\sum_{a<t_i<b} \varepsilon_j(t_i) \leq c_n \sigma^2(b-a) \eqsp, $ where $c_n\to0$ and $\lim_{\delta\to 0} \sigma^2(\delta)= 0$. One important situation in which the error terms are not \iid\  is when an adaptive sampling strategy is adopted, such that the acquisition of the observations is concentrated around interesting points of the signal. This discussion is beyond the scope of this paper.

\subsection{Heuristic argument and asymptotic expansion}

Before detailing the complete derivation of the estimate properties, we give in this section a heuristic argument which shows the consistency and of $\boldsymbol\theta_1$ in a simple case. We assume for simplicity that  \(M = K\gg n\to\infty\) and that all the shifts $\{\theta_j,\ j=0\ldots M\}$ are equal to zero, so that $\boldsymbol\alpha_1$ represents only the error made during alignment, and we only have one block to process. We also assume that the signal \(s\) is an odd function, so that \(c_s(k)= \mathrm{i} c_k\) is a non-zero imaginary number, and there is no reason to align the curves with respect to $y_0$. We assume,  without any loss of generality, that  $\sum_{m=0}^K\alpha_m=0$. Define for all integers $k$ and $l$ : \(\tilde V_{k,l} \eqdef \sigma n^{-1/2}V_{k,l}\) and  \(\tilde W_{k,l} \eqdef \sigma n^{-1/2}W_{k,l}\), and define the random variables $R_{k,l}$ and $\beta_{k,l}$ so that we can write $\tilde{V}_{k,l} + \rmi\tilde{W}_{k,l}=R_{k,l} \rme^{\rmi\beta_{k,l}}$.  Observe now that since
\begin{equation}
C_1(k,\boldsymbol\alpha_1) = \frac{1}{K+1} \sum_{l=0}^K |f_{k,l}|^2 - \left|\frac{1}{K+1} \sum_{l=0}^K \rme^{\rmi \alpha_l k}f_{k,l}\right|^2 \eqsp ,
\label{eq:specific_form_C1}
\end{equation}
then $ C_1(k,\boldsymbol\alpha_1) \geq 0$ due to Cauchy-Schwarz inequality. Since $K$ tends to infinity, the mean energy spectral density, the first term on the right-hand-side (RHS) of \eqref{eq:specific_form_C1} is approximately \(c_k^2 + O_\PP (K^{-1/2})\), so that:
\begin{align*}
&C_1(k,\boldsymbol\alpha_1)  = c_k^2 + \mathrm O_\PP (K^{-1/2}) \\& ~~~~~~ - \left|  \frac{1}{K+1} \sum_{l=0}^K e^{i\alpha_lk}(\tilde V_{k,l} + \mathrm{i} (c_k + \tilde W_{k,l})) \right |^2 \\
%&= c_k^2 + \mathrm O_\PP (M^{-1/2}) \\
%& ~~~- \Bigl(\frac{1}{K+1} \sum_{l=0}^K \bigl(\tilde V_l\cos(\alpha_l k) -(c_k+ \tilde W_l)\sin(\alpha_l k)\bigr)\Bigr)^2 \\
%& ~~~~~- \Bigl(\frac{1}{K+1} \sum_{l=0}^K \bigl(\tilde V_l\sin(\alpha_l k) +(c_k+ \tilde W_l)\cos(\alpha_l k)\bigr)\Bigr)^2,\\
%&= c_k^2 +  \mathrm O_\PP (K^{-1/2}) \\
%& ~~~- \Bigl(\frac{1}{K+1} \sum_{l=0}^K \bigl(R_{k,l}\cos(\alpha_l k+\beta_{k,l}) -c_k\sin(\alpha_l k)\bigr)\Bigr)^2 \\
%& ~~~~~- \Bigl(\frac{1}{K+1} \sum_{l=0}^K \bigl(R_{k,l}\sin(\alpha_l k+\beta_{k,l}) +c_k\cos(\alpha_l k)\bigr)\Bigr)^2
%\\
&= c_k^2 + \mathrm O_\PP (K^{-1/2})
\\& - \frac{1}{(K+1)^2}\sum_{l=0}^K\sum_{m=0}^K \Bigl( c_k^2 \cos((\alpha_l-\alpha_m)k)
\\&- 2 c_k R_{k,m} \sin((\alpha_l-\alpha_m)k-\beta_{k,m})
%-  c_k R_{k,l} \sin((\alpha_m-\alpha_l)k-\beta_{k,l})
\\&+ R_{k,l}R_{k,m} \cos((\alpha_l-\alpha_m)k+\beta_{k,l}-\beta_{k,m}) \Bigr)   \eqsp .
\end{align*}
Expanding the harmonic functions up to $\mathrm{o}_\PP (n^{-1})$, assuming that the  $\alpha_m=\mathrm O_\PP(n^{-1/2})$  (which shall be later proved in Theorem \ref{th:markov_like}), and noting that $R_{k,m}=\mathrm O_\PP(n^{-1/2})$, we get for any fixed $k$ that
\begin{equation*}
\begin{split}
& \frac{-1}{(K+1)^2} \sum_{l=0}^K \sum_{m=0}^K R_{k,l}R_{k,m}\cos((\alpha_l-\alpha_m)k+\beta_{k,l}-\beta_{k,m}) \\
&= - \frac{1}{(K+1)^2} \sum_{l=0}^K \sum_{m=0}^K R_{k,l}R_{k,m}\cos(\beta_{k,l}-\beta_{k,m}) \\
&+\frac{1}{(K+1)^2} \sum_{l=0}^K \sum_{m=0}^K R_{k,l}R_{k,m}\sin(\beta_{k,l}-\beta_{k,m})(\alpha_l-\alpha_m) k\\&
~~~~~~~~~+ \mathrm o_\PP(n^{-1})
\\
&= - \frac{1}{(K+1)^2} \sum_{l=0}^K \sum_{m=0}^K R_{k,l}R_{k,m}\cos(\beta_{k,l}-\beta_{k,m}) \\
&~~~~~~~~~ + \mathrm o_\PP(n^{-1}) \eqsp .
\end{split}
\end{equation*}
Since we assumed that $\sum_{m=0}^K \alpha_m=0$, we obtain:
\begin{equation*}
\begin{split}
& \frac{1}{(K+1)^2} \sum_{l=0}^K\sum_{m=0}^K 2 c_k R_{k,m} \sin((\alpha_l-\alpha_m)k-\beta_{k,m}) \\
&=  - \frac{2c_k}{K+1}\sum_{m=0}^K R_{k,m}\sin(\beta_{k,m})\\
& + \frac{2kc_k}{(K+1)^2}\sum_{m=0}^K R_{k,m} \cos(\beta_{k,m})\sum_{l=0}^K(\alpha_l-\alpha_m)+ \mathrm o_\PP(n^{-1})
\\
&=  -\frac{2c_k}{K+1}\sum_{m=0}^K R_{k,m}\sin(\beta_{k,m}) \\
&- \frac{2kc_k}{(K+1)^2}\sum_{m=0}^KR_{k,m} \cos(\beta_{k,m})\alpha_m+ \mathrm o_\PP(n^{-1}) \eqsp .
\end{split}
\end{equation*}
Using the same assumption, we get that
\begin{equation*}
\begin{split}
& - \frac{1}{(K+1)^2}\sum_{l=0}^K\sum_{m=0}^K c_k^2 \cos((\alpha_l-\alpha_m)k) \\
&=  -c_k^2 + \frac{k^2c_k^2}{2(K+1)^2}\sum_{l=0}^K\sum_{m=0}^K(\alpha_l-\alpha_m)^2 + \mathrm o_\PP(n^{-1})
\\
&=  -c_k^2 + \frac{k^2c_k^2}{K+1}\sum_{l=0}^K\alpha_l^2 + \mathrm o_\PP(n^{-1}) \eqsp .
\end{split}
\end{equation*}
Hence, the Taylor expansion of the cost function is equal to

\begin{equation*}
\begin{split}
&C_1(k,\boldsymbol\alpha_1) \\
&=   -\frac{1}{(K+1)^2} \sum_{l=0}^K \sum_{m=0}^K R_{k,l}R_{k,m}\cos(\beta_{k,l}-\beta_{k,m})
\\
& - \frac{2kc_k}{K+1}\sum_{m=0}^KR_{k,m} \cos(\beta_{k,m}) \alpha_m \\&-\frac{2c_k}{K+1}\sum_{m=0}^K R_{k,m}\sin(\beta_{k,m})
\\
&+\frac{k^2c_k^2}{K+1}\sum_{m=0}^K\alpha_m^2 +\mathrm O_\PP(K^{-1/2})+\mathrm o_\PP(n^{-1}) \eqsp,
\end{split}
\end{equation*}
which is minimized by taking $\hat\theta_m = R_{k,m}\cos(\beta_{k,m})/kc_k+\mathrm o_\PP(n^{-1/2}) + \mathrm{O}_\PP (K^{-1/2})$.
More generally, when the different bands are weighted, we obtain by differentiation
\begin{equation*}
\displaystyle \hat\theta_m = \frac{\sum_{k\in{\cal K}}\nu_k kc_k^3 R_{k,m}\cos(\beta_{k,m})}{\sum_{k\in{\cal K}}\nu_k k^2  c_k^4}+\mathrm o_\PP(n^{-1/2}) + \mathrm{O}_\PP (K^{-1/2})
\end{equation*}
which establishes the asymptotic expansion (up to the first order) and the asymptotic normality of the estimate when both $n$ and $K$ tend to infinity.

\subsection{Computation of the cost function $C_m$ }

The cost function $C_m$ associated with block $m$ can be written as follows:
\begin {equation}
\begin{split}
&\hspace{-2em}\|C_m(\boldsymbol{\alpha}_m)\|_\nu^2
%\\&
= \sum_{k\in{\cal K}}\nu_k\left(A_M(k)-B_m(k,\boldsymbol{\theta}_m)\right)^2 \\
&+
\sum_{k\in{\cal K}}\nu_k\left(B_m(k,\boldsymbol{\theta}_m)-B_m(k,\boldsymbol{\alpha}_m)\right)^2
 \\
&
+2\sum_{k\in{\cal K}}\nu_k\left(B_m(k,\boldsymbol{\theta}_m)-B_m(k,\boldsymbol{\alpha}_m)\right)
\label{eq:exact_approx}
\\
&\hspace{5em}\times\left(A_M(k)-B_m(k,\boldsymbol{\theta}_m)\right),
% \\
% &+ \sum_{k=0}^{n-1}\left(B_m(k,\boldsymbol{\theta}_m)-B_m(k,\boldsymbol{\alpha}_m)\right)\left(A_M(
% k)-B_m(k,\boldsymbol{\theta}_m)\right)^{*} \label{eq:crossed_term2}
\end{split}
\end{equation}
where $A_M(k)$ and $B_m(k,\boldsymbol{\alpha}_m)$ are the first and second terms of the RHS of~\eqref{eq:aux_function}, both taken at point $k$. We focus on the expansion of the terms associated with $\|C_1(\boldsymbol{\alpha}_1)\|_\nu^2$, since all other cost functions may be expanded in a similar manner up to a change of index. We detail the expansion of $A_M(k)$ and $B_1(k,\boldsymbol{\alpha}_1)$, since $B_1(k,\boldsymbol{\theta}_1)$ can be easily obtained from the latter term.

Recall that $A_M(k) = \frac{1}{M+1} \sum_{l=0}^M |f_{k,l}|^2$; we get that
\begin{align*}
& A_M(k) =  \frac{1}{M+1} \sum_{l=0}^M \left|e^{-ik\theta_l} c_s(k) + \frac{\sigma}{\sqrt{n}} (V_{k,l} + i W_{k,l}) \right|^2 \\
%&=  \frac{1}{M+1} \sum_{l=0}^M \left(e^{-ik\theta_l} c_s(k) + \frac{\sigma}{\sqrt{n}} (V_{k,l} + i W_{k,l})  \right) \\ & \times\left(e^{ik\theta_l} c_s^{*}(k) + \frac{\sigma}{\sqrt{n}} (V_{k,l} - i W_{k,l}) \right) \\
&= \frac{1}{M+1} \sum_{l=0}^M \Bigl\{ |c_s(k)|^2 + \frac{\sigma^2}{n} (V_{k,l}^2 + W_{k,l}^2) \\
& + \frac{2\sigma}{\sqrt{n}}V_{k,l} \mathrm{Re}(e^{-ik\theta_l}c_s(k))  + \frac{2 \sigma}{\sqrt{n}}W_{k,l} \mathrm{Im}(e^{-ik\theta_l}c_s(k))   \Bigr\}
\end{align*}
Due to the equalities $\mathrm{Re}(e^{-k\theta_l}c_s(k)) = \cos(k\theta_l) \mathrm{Re}(c_s(k)) + \sin(k\theta_l) \mathrm{Im}(c_s(k))  $ and $\mathrm{Im}(e^{-k\theta_l}c_s(k)) = \cos(k\theta_l) \mathrm{Im}(c_s(k)) - \sin(k\theta_l) \mathrm{Re}(c_s(k))  $, it follows that

\begin{align}
& A_M(k) = |c_s(k)|^2 +  \frac{\sigma^2}{n(M+1)} \sum_{l=0}^M (V_{k,l}^2+W_{k,l}^2) \label{eq:computation_A} \\
&+ \frac{2\sigma\mathrm{Re}(c_s(k))}{\sqrt{n}(M+1)} \sum_{l=0}^M \left( V_{k,l} \cos(k\theta_l) - W_{k,l} \sin(k\theta_l)\right) \nonumber \\
& + \frac{2\sigma\mathrm{Im}(c_s(k))}{\sqrt{n}(M+1)} \sum_{l=0}^M
 \left( V_{k,l} \sin(k\theta_l) + W_{k,l} \cos(k\theta_l) \right) \nonumber
\end{align}

%\end{split}\end{equation}
%
\begin{remark} \label{rem:convergence_A}
 By Assumption~\ref{hyp:independence} and the law of large numbers the last two terms of~\eqref{eq:computation_A} converge almost surely to $0$ as $M$ tends to infinity. Moreover, the sum of the second term has  a $\chi^2$ distribution with $2(M+1)$ degrees of freedom. Thus, the term $A_M(k)$ tends to $\left|c_s(k) \right|^2 + 4n^{-1}\sigma^2$ as $M\to\infty$, and therefore to $S_s (k)$  as both $M$ and $n$ tend to infinity. %$\left|c_s(k) \right|^2$ as both $M$ and $n$ tend to infinity.
\end{remark}
The first curve of each block is the reference curve, which is
considered to be invariant and thus has a known associated shift, so
that $\alpha_0=\theta_0=\hat\theta_0=0$. It stems from
\eqref{eq:reshifted} and \eqref{eq:discrete_model_fourier} that
% Recall that $B_m(k,\boldsymbol{\alpha}_m)$ is the modulus of the squared DFT of the average of the curves in block $m$, after shift correction. %  Up to a change of index, % it is possible to write that each curve $l$ of block % $j$ has a shift $\theta_l$ and an associated correction term of $\alpha_{l}$. The first curve of each block is the reference curve, which is considered to be invariant and thus has a known associated shift $\alpha_0=\theta_0=0$. We obtain
\begin{equation*}\begin{split}
&B_1(k,\boldsymbol{\alpha}_1)
%\\ &
= \left| \frac{1}{\lambda+K}\left[\lambda
    ( c_s(k) + \frac{\sigma}{\sqrt{n}} (V_{k,0}+\rmi
      W_{k,0})) \right.\right.
      \\
      & \left.\left. + \sum_{l=1}^{K} \left( \rme^{\rmi k
        (\alpha_l-\theta_l)} c_s(k) + \frac{\sigma}{\sqrt{n}}
      \rme^{\rmi k \alpha_l} (V_{k,l}+\rmi W_{k,l})\right)
  \right]\right|^2 \eqsp,
\end{split}\end{equation*}
thus, if we define  $\lambda_m$, $m=0\ldots K$, such that
$\lambda_0 \eqdef \lambda$ and $\lambda_m \eqdef 1$ otherwise:
\begin{equation*}
\begin{split}
&B_1(k,\boldsymbol{\alpha}_1)
= \frac{1}{(K+\lambda)^2} \\& \times\left( \sum_{l=0}^K \lambda_l \left( \rme^{\rmi k (\alpha_l-\theta_l)} c_s(k) + \frac{\sigma}{\sqrt{n}} \rme^{\rmi k \alpha_l} (V_{k,l}+\rmi W_{k,l}) \right) \right) \\
      &\times \left(\sum_{m=0}^K \lambda_m \left(  \rme^{\rmi k (\theta_m-\alpha_m)} c_s^{*}(k) + \frac{\sigma}{\sqrt{n}} \rme^{-\rmi k \alpha_m} (V_{k,m}-\rmi W_{k,m})  \right) \right) \eqsp,
\end{split}
\end{equation*}
and expanding the latter yields
%
%\begin{equation}\begin{split}
\begin{align}
&B_1(k,\boldsymbol\alpha_1)
\nonumber
%\\ &
=\frac{|c_s(k)|^2}{(\lambda+K)^2}\sum_{l,m=0}^K \lambda_l \lambda_m
  \rme^{\rmi k(\alpha_l-\theta_l-\alpha_m+\theta_m)}
\nonumber  \\
 &+ \frac{\sigma^2}{n(\lambda+K)^2}\sum_{l,m=0}^K \lambda_l
 \lambda_m \{\rme^{\rmi k (\alpha_l-\alpha_m)} \times
 \label{eq:computation_Bjalpha}\\
&  [V_{k,l}V_{k,m}
 + W_{k,l}W_{k,m} +\rmi (V_{k,l}W_{k,m}
 - W_{k,l}V_{k,m})]\}
 \nonumber\\
 &+\frac{\sigma c_s(k)}{\sqrt{n}(\lambda+K)^2}\sum_{l,m=0}^K \lambda_l \lambda_m
\rme^{\rmi k (\alpha_l-\theta_l-\alpha_m)}(V_{k,m}-\rmi W_{k,m})
\nonumber\\
 &+\frac{\sigma c_s^{*} (k)}{\sqrt{n}(\lambda+K)^2}\sum_{l,m=0}^K \lambda_l \lambda_m
 \rme^{\rmi k (\theta_m+\alpha_l-\alpha_m)} (V_{k,l}+\rmi W_{k,l}) \eqsp .
 \nonumber
  \end{align}
%\end {split}\end{equation}
%
The functional $\|C_1(\boldsymbol\alpha_1)\|_\nu^2$ can be split into a stochastic part which depends on the random variables $\left\{V_{k,l}, \  k=-\frac{n-1}{2} \ldots \frac{n-1}{2} \right\}$ and $\left\{W_{k,l}, \  k= -\frac{n-1}{2} \ldots \frac{n-1}{2}\right\}$, and a noise-free part which does not depends on them, and is further on denoted by $D_1(\boldsymbol\alpha_1)$. Observe that the first sum in \eqref{eq:computation_Bjalpha} is equal to $|c_s(k)|^2$ when taking $\boldsymbol\alpha_1=\boldsymbol\theta_1$; consequently, all the terms stemming from the first and the third sum in \eqref{eq:exact_approx} depend on the random variables $\left\{V_{k,l}, \  k=-\frac{n-1}{2} \ldots \frac{n-1}{2} \right\}$ and $\left\{W_{k,l}, \  k= -\frac{n-1}{2} \ldots \frac{n-1}{2}\right\}$, and are not part of the functional $D_1(\boldsymbol\alpha_1)$. This term is equal to:
\begin{align}    \label{eq:deterministic_part}
&D_1(\boldsymbol\alpha_1)
 \\
&= \sum_{k\in \cal K}\nu_k |c_s(k)|^4 \left| \left|\frac{1}{K+\lambda} \sum_{m=0}^K
    \lambda_m \rme^{\rmi k (\alpha_m - \theta_m)}\right|^2 - 1\right|^2
    \nonumber
\end{align}
Details of the calculations are given in Appendix \ref{sec:noise-free-computation}. Note that due to~\eqref{eq:deterministic_part}, $D_1$ has a unique global minimum which is attained when $\alpha_m = \theta_m$, for all $m=1\dots,K$, that is the actual shift value. We show in Proposition~\ref{prop:noisy-part} that $\|C_1(\boldsymbol\alpha_1)\|_\nu^2-D_1(\boldsymbol\alpha_1)$ is negligible when both $n$ and $K$ tend to infinity, under mild assumptions on $\lambda$, so that the proposed cost function behaves asymptotically like $D_1(\boldsymbol\alpha_1)$.
\begin{proposition}
\label{prop:noisy-part}
Assume that $K\to\infty$, $n\to\infty$, $\lambda\to\infty$, and $\lambda/K \to 0$. Denote the noise-free part associated with $B_1(k,\boldsymbol\theta_1) - B_1(k,\boldsymbol\alpha_1)$ by $\Delta(k,\boldsymbol\alpha_1)$, that is
\begin{equation*}
\Delta(k,\boldsymbol\alpha_1) \eqdef  |c_s(k)|^2 \left| \left|\frac{1}{K+\lambda} \sum_{m=0}^K \lambda_m \rme^{\rmi k (\alpha_m - \theta_m)}\right|^2 - 1\right| ,
\end{equation*}
And denote the noise part by $R(k,\alpha_1)\eqdef B_1(k,\boldsymbol{\theta}_1) - B_1(k,\boldsymbol{\alpha}_1) - \Delta (k,\boldsymbol\alpha_1)$.
Then:
\begin{align}
\label{sumsqdiff}
&\sum_{k\in \cal K} \nu_k\bigl(A_M(k)-B_1(k,\boldsymbol{\theta}_1)
\bigl)^2
=  \mathrm O_\PP \left(\frac{1}{n^2}\right) + \mathrm O_\PP \left(\frac{1}{nK}\right) \nonumber
\\
&\sum_{k\in \cal K} \nu_k R(k;\boldsymbol\alpha_1)^2
= \mathrm O_\PP \left(\frac{1}{n^2} \right) \nonumber
\\
&+  \mathrm O_\PP \Bigl(\frac {1} { n} \Bigr)\inf_c\frac1{K+\lambda}
\sum_{m=0}^K\lambda_m(\alpha_m-\theta_m-c)^2  
\\
&\|C_1(\boldsymbol{\alpha}_1)\|_\nu^2 = \sum_{k\in \cal K}\nu_k \Delta(k, \boldsymbol\alpha_1)^2+  \mathrm O_\PP \left(\frac{1}{nK}\right)+\mathrm O_\PP \left(\frac{1}{n^2} \right) \nonumber
\\
&+  \biggl(\mathrm O_\PP \Bigl(\frac {1} { n} \Bigr)+\mathrm O_\PP \Bigl(\frac {1} { \sqrt{nK}} \Bigr)\biggr)\inf_c\frac1{K+\lambda}
\sum_{m=0}^K\lambda_m(\alpha_m-\theta_m-c)^2 \nonumber \\
& +  \mathrm O_\PP \Bigl(\frac {1} {\sqrt{ n}} \Bigr)\left[\inf_c\frac1{K+\lambda}
\sum_{m=0}^K\lambda_m(\alpha_m-\theta_m-c)^2\right]^{3/2}  \nonumber.
\end{align}
where the $\mathrm O_\PP$ hold uniformly in $\boldsymbol \alpha_1$. \nonumber
\end{proposition}
\begin{IEEEproof}
See Appendix~\ref{sec:proof-noisy-part}.
\end{IEEEproof}

%We conclude from Proposition~\ref{prop:noisy-part}  that if each block contains a large number of curves, and  the weighting factor $\lambda$ is large but negligible relative to $K$,   the cost function $\|C_1(\boldsymbol\alpha_1)\|_\nu^2$ converges to $D_1(\boldsymbol\alpha_1)$.
Since \(\boldsymbol{\hat\theta}_1\) is the minimizer of \(C_1(\boldsymbol\alpha_1) \) and \(D_1(\boldsymbol\theta_1)=0\), we get by means of Proposition~\ref{prop:noisy-part} that
\begin{equation}
\label{dthsmall}
\begin{split}
& D_1(\boldsymbol{\hat\theta}_1)
 = \|C_1(\boldsymbol{\hat\theta}_1)\|_\nu^2  +\bigl(D_1(\boldsymbol{\hat\theta}_1)- \|C_1(\boldsymbol{\hat\theta}_1)\|_\nu^2 \bigr)
 \\
& \leq \|C_1(\boldsymbol\theta_1)\|_\nu^2  + \bigl(D_1(\boldsymbol{\hat\theta}_1)- \|C_1(\boldsymbol{\hat\theta}_1)\|_\nu^2 \bigr)
 \\
& = D_1(\boldsymbol\theta_1) + \bigl(D_1(\boldsymbol{\hat\theta}_1)-\|C_1(\boldsymbol{\hat\theta}_1)\|_\nu^2 \bigr)
\\&\hspace{3em}
 - \bigl(D_1(\boldsymbol\theta_1) -  \|C_1(\boldsymbol\theta_1)\|_\nu^2  \bigr)
 \\
& =\bigl(D_1(\boldsymbol{\hat\theta}_1)- \|C_1(\boldsymbol{\hat\theta}_1) \|_\nu^2\bigr)
\\&\hspace{3em}
 -\bigl (D_1(\boldsymbol\theta_1) -  \|C_1(\boldsymbol\theta_1)\|_\nu^2  \bigr)
\\
&=\mathrm O_\PP \left(\frac{1}{n^2} \right) + \mathrm O_\PP \left(\frac{1}{nK}\right)
\\
&+   \biggl(\mathrm O_\PP \Bigl(\frac {1} { n} \Bigr)+\mathrm O_\PP \Bigl(\frac {1} { \sqrt{nK}} \Bigr) \biggr) \inf_c \frac 1 {K+\lambda} \sum_{m=0}^K\lambda_m(\hat\theta_m-\theta_m-c)^2 \\
&+ \mathrm O_\PP \Bigl(\frac {1} {\sqrt{n}} \Bigr)\left[ \inf_c \frac 1 {K+\lambda} \sum_{m=0}^K\lambda_m(\hat\theta_m-\theta_m-c)^2 \right]^{3/2} \eqsp ,
\end{split}
\end{equation}
thus showing that $D_1(\boldsymbol{\hat\theta}_1)$ is close to zero as both $n$ and $K$ tend to infinity. The main result is using the fact that the only minimizer of $D_1$ is the true vector of shifts.

\subsection{Theoretical properties of the shift estimation algorithm}

The following result gives information on the number of curves well aligned in a given block, and holds for each term in the sum of Equation~(\ref{eq:deterministic_part}).

\begin{proposition}
\label{prop:bad_control_deterministic}
Let $\eta \to 0$ as $n,K\to\infty$, $\lambda \geq 1$ and let $\delta$ be a real positive number. Suppose that $\alpha_1,\dots,\alpha_m$ is any sequence  such that:
$$
\left| \frac{1}{(K+\lambda)} \sum_{m=0}^K \lambda_m e^{\rmi k  \, (\theta_m-\alpha_m) }\right| >
1-\eta \eqsp,
$$
for some \(k \in \cal K\). Then there exist two positive constants $\gamma_0$ and $K_0$, such that for $K\geq K_0$, there is a constant $c$ such that the number of curves whose alignment error
$\alpha_m-\theta_m-c$ is bigger than $\eta^\delta$,  is bounded by $
 \gamma_0 (K+\lambda) \eta^{1-2\delta} $.  Moreover,
\begin{align}\label{sqErr}
\sum_{m=1}^K(\theta_m-\alpha_m-c)^2\leq\
\frac{(K+\lambda)\eta}{\gamma_0k^2}.
\end{align}
\end{proposition}

\begin{IEEEproof}
See Appendix~\ref{sec:proof_bad_control_deterministic}.
\end{IEEEproof}

Note that the latter proposition is of interest only when $ 0 < \delta < 1/2$, since $\delta > 1/2$ would yield a large upper bound. Proposition~\ref{prop:bad_control_deterministic}  has the following motivation: when the number of curves in each block is large enough, the noise contribution to the criterion will be small, and \(\boldsymbol{\hat\theta}_1\) will be such that the condition of the proposition holds. Hence, we can conclude that most curves will tend to align. However, they may not align with the reference curve $y_0$. Consequently, the weighting factor $\lambda$ is introduced in order to ``force'' all the curves in a block to align with respect to $y_0$, as stated in the following proposition:
\begin{proposition}
\label{prop:choice_lambda}
Assume that $\lambda$ is an integer, and that \(\eta^{1-2\delta}\leq \lambda/(\gamma_0(K+\lambda))\), where $\gamma_0$ is the positive constant appearing in the previous proposition. Then, under the assumption of Proposition~\ref{prop:bad_control_deterministic}, we get that $|c| < \eta^{\delta}$.
\end{proposition}

\begin{IEEEproof}
See Appendix~\ref{sec:proof_choice_lambda}.
\end{IEEEproof}

In other words, when  $\lambda$ is chosen such that $\lambda\to \infty$ and ${\lambda}/{K}\to 0$ as $K\to\infty$, the estimate will be close to the actual shifts. We now state the main theorem:

\begin{theorem}
\label{th:markov_like}
Under Assumptions \ref{hyp:finite_support}--\ref{hyp:ind}, if \(K\to\infty\), $n\to \infty$, $\lambda = \lambda(K) \to\infty$,   $n^{1/4}\lambda/K\to0$,  and \(n/K\) is bounded,  then for all  \(\delta\in(0,1/2)\), there exists \(\gamma>0\),  such that with probability converging to 1
\begin{equation}
\label{th31result}
\begin{split}
\frac 1{K+\lambda}\sum_{m=0}^K \boldsymbol 1(|\hat\theta_m-\theta_m|> 2 n^{-\delta})&\leq \gamma n^{-(1-2\delta)}.
\end{split}
\end{equation}

\end{theorem}
\begin{proof}
In the following, $\gamma_1,\gamma_2,\ldots$ denote  positive constants such that the corresponding inequalities hold. The proof of this theorem is  deduced from \eqref{dthsmall} and Propositions~\ref{prop:bad_control_deterministic} and \ref{prop:choice_lambda}. 

Define
 \begin{align*}
 A^2 \eqdef \inf_c\frac1{K+\lambda}\sum_{m=0}^K(\hat\theta_m-\theta_m-c)^2.
 \end{align*}
By \eqref{eq:deterministic_part} and \eqref{dthsmall} we can use Proposition \ref{prop:bad_control_deterministic} with
\begin{equation}
\label{etadef}
\eta= \frac{\gamma_1} n+\frac {\gamma_1} {\sqrt n}A + \frac {\gamma_1} {n^{1/4}}A^{3/2} ,
\end{equation}
Since $\theta_m$ and $\hat\theta_m$ are bounded, we obtain that $\eta=\mathrm o_\PP(1)$. Equation~\eqref{sqErr} in Proposition \ref{prop:bad_control_deterministic} yields
\begin{align*}
& A^2
\leq
\frac{\gamma_1}{n}+\frac{\gamma_2}{\sqrt n}A + \frac{\gamma_3}{n^{1/4}}A^{3/2}
\end{align*}
Define $B\eqdef \sqrt{n}A$, so that the latter becomes $B^2 - \gamma_1 - \gamma_2 B - \gamma_3 B^{3/2} \leq 0$. A continuity argument yields the boundedness of $B$, thus:
\begin{align}\label{sqErr2}
\inf_c\frac1{K+\lambda}\sum_{m=0}^K(\hat\theta_m-\theta_m-c)^2
 \leq \frac{\gamma_4}n \eqsp,
\end{align}
which shows that $\eta \leq \gamma_5 /n$. On the other hand, by \eqref{eq:deterministic_part}, \eqref{dthsmall}, and Proposition~\ref{prop:choice_lambda} we conclude that with probability converging to 1:
\begin{equation}
\label{Mlike}
\begin{split}
\frac 1{K+\lambda}\sum_{m=0}^K \boldsymbol 1(|\hat\theta_m-\theta_m|> 2 \eta^{\delta})&\leq \gamma_6 \eta^{1-2\delta}\eqsp,
\end{split}
\end{equation}
%Using the boundedness of $\hat\theta_m$ and $\theta_m$ we obtain from \eqref{Mlike} with $\delta = 1/4$ that  with probability converging to 1:
%\begin{equation}
%\label{Mlike2}
%\begin{split}
%& \frac 1{K+\lambda}\sum_{m=0}^K (\hat\theta_m-\theta_m)^2 \\
%& = \frac 1{K+\lambda}\sum_{m=0}^K  (\hat\theta_m-\theta_m)^2 (\boldsymbol1(|\hat\theta_m-\theta_m|> 2 \eta^{1/4}) + \boldsymbol1(|\hat\theta_m-\theta_m|\leq 2 \eta^{1/4})) \\
%& \leq ( (2\pi)^2 \gamma_2 \eta^{1-2/4} + 2\eta^{1/2} ) \leq \gamma_3 \eta^{1/2} \eqsp.
%\end{split}
%\end{equation}
%
and due to \eqref{sqErr2}, \eqref{Mlike} still holds when replacing $\eta$ by $\gamma_5/n$, thus proving \eqref{th31result} and the theorem. In particular, letting $\delta$ be as close to $1/2$ as needed shows that the estimator $\boldsymbol{\hat\theta}_1$ tends to $\boldsymbol\theta_1$ with the standard rate of convergence $n^{-1/2}$.
\end{proof}

\subsection{Weak convergence of the density estimator}

Due to the previous results, it is now possible to give a theoretical result about the plug-in estimate of the distribution of $\theta$. As suggested in~\eqref{eq:kernel_estimate}, an estimate of the probability density function $f$ can be obtained by plugging the approximated values of the shifts into a known density estimate. We provide here a result on the weak convergence of the empirical estimator.
\begin{theorem}
\label{prop:weak_convergence}
Let $g$ be a continuous function with a bounded derivative. Under the assumptions of Theorem~\ref{th:markov_like}, we get almost surely when $M \to \infty, n \to \infty$ that
\begin{equation}
\frac{1}{M+1} \sum_{m=0}^M g(\hat{\theta}_m) \longrightarrow
\E[g(\theta)]. \label{eq:weak_convergence}
\end{equation}
\end{theorem}
Proof of theorem~\ref{prop:weak_convergence} can be sketched as follows: due to the Law of Large Numbers, it is equivalent to show that:
$$
\frac{1}{M+1} \sum_{m=0}^M (g(\hat\theta_m) - g(\theta_m))
$$
converges almost surely to 0. Since $g$ has a bounded derivative, we can write that the absolute value of the latter term is bounded by
$$
\frac{\sup_x |g'(x)|}{M+1} \sum_{m=0}^M |\hat\theta_m-\theta_m|.
$$
Consequently, due to Theorem~\ref{th:markov_like}, there exists a constant C such that with probability:
$$
\frac{1}{M+1} \sum_{m=0}^M (g(\hat\theta_m) - g(\theta_m)) \leq C \left(\frac{1}{n^{\delta}} + \frac{1}{n^{(1-2\delta)}} \right),
$$
which completes the proof. More particularly, taking $g(\cdot) = h^{-1}\psi\left(\frac{\cdot - x}{h}\right)$, where $h^2\min\{n^\delta,n^{1-\delta}\}\to\infty$ we get that~\eqref{eq:kernel_estimate} tends to $\E\left[h^{-1} \psi\left(\frac{\theta-x}{h}\right)\right]$, thus showing pointwise consistency, that is
$$
\hat{f}_{M,h} (x) \longrightarrow f(x) \text{ as } M\to\infty,h\to 0\ , Mh \to \infty \,
$$
for any continuity point $x$ of $f$.
\begin{remark}
If \(n\) remains bounded as \(K\to \infty\), then the parameters \(\theta_m\) cannot be estimated without an error, and the observed distribution of \(\{\hat\theta_m\}\) would be a convolution of the distribution of \(\{\theta_m\}\) with the estimation error. If \(n\) is large enough, the latter distribution is approximately normal with variance which is \(\mathrm O_\PP(\sigma^2/n)\).
\end{remark}

\begin{remark}
The discussion was under the assumption that the $\theta_1,\dots,\theta_M$ have a continuous distribution with a smooth density.  If this would not be the case, then the estimated density will be approximately equal to a smoothed version of the distribution.
\end{remark}

\section{Applications}

\begin{figure*}[thp]
\centering
  \subfigure[]{\includegraphics[width=0.32\linewidth]{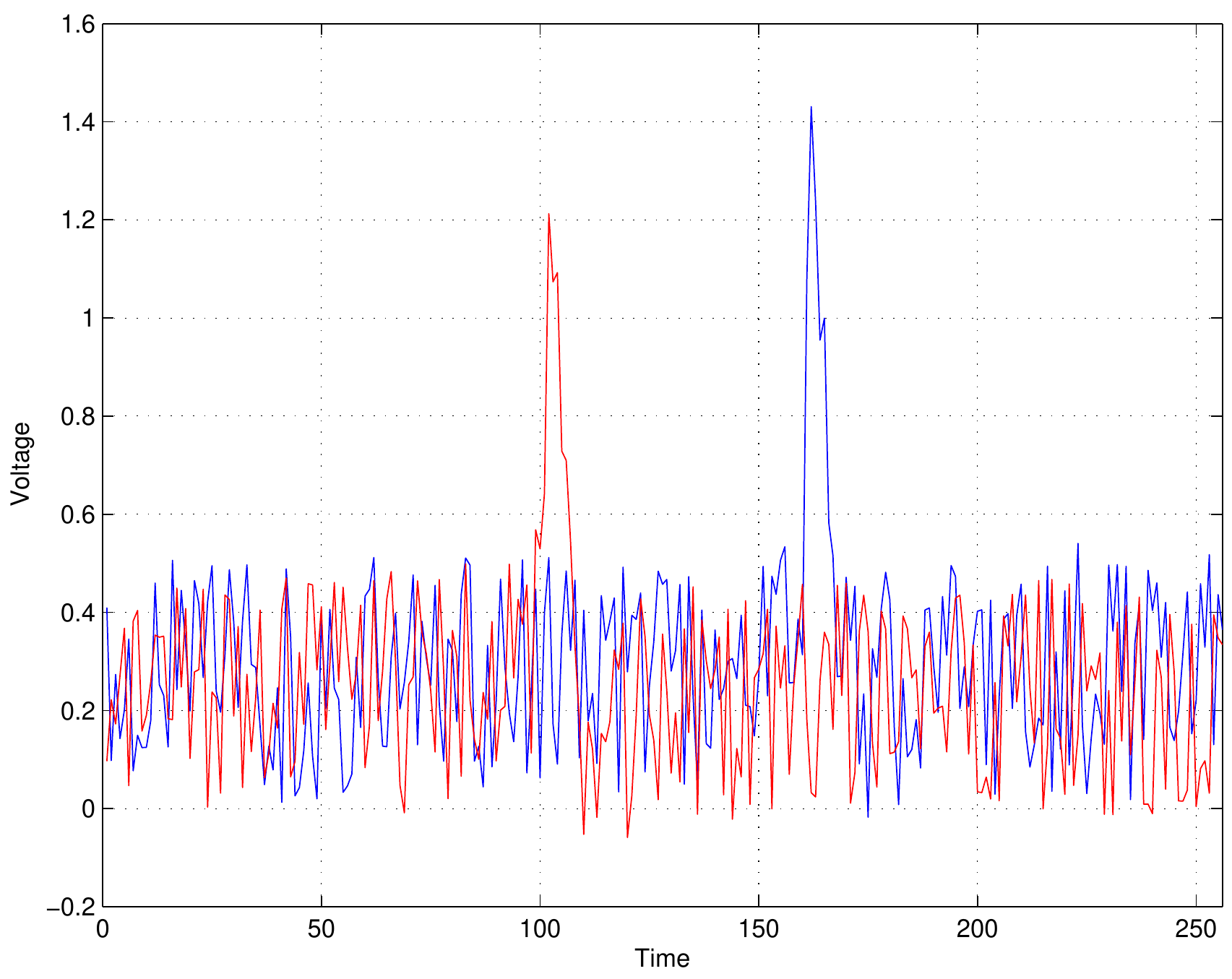}}
  \subfigure[]{\includegraphics[width=0.32\linewidth]{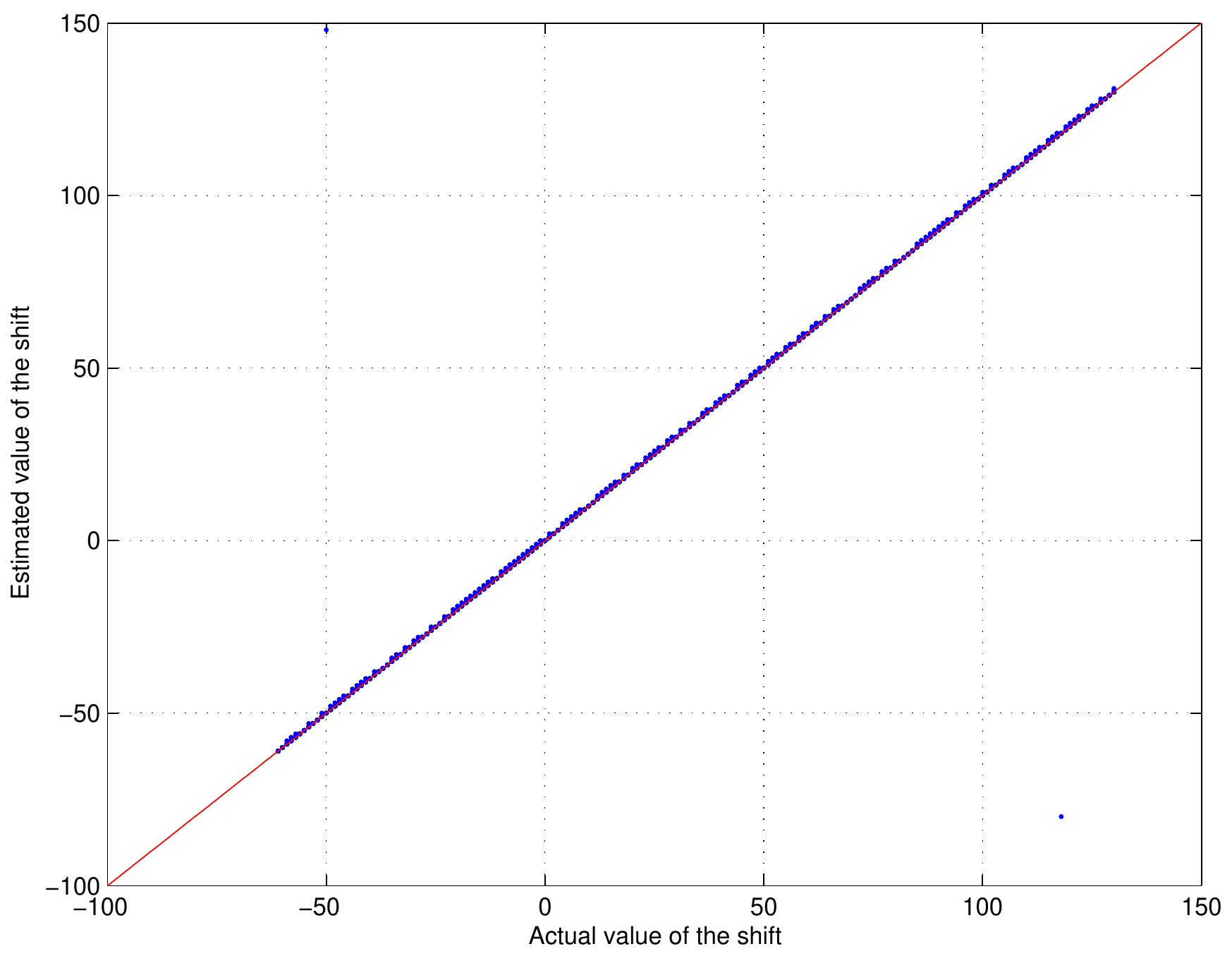}}
  \subfigure[]{\includegraphics[width=0.32\linewidth]{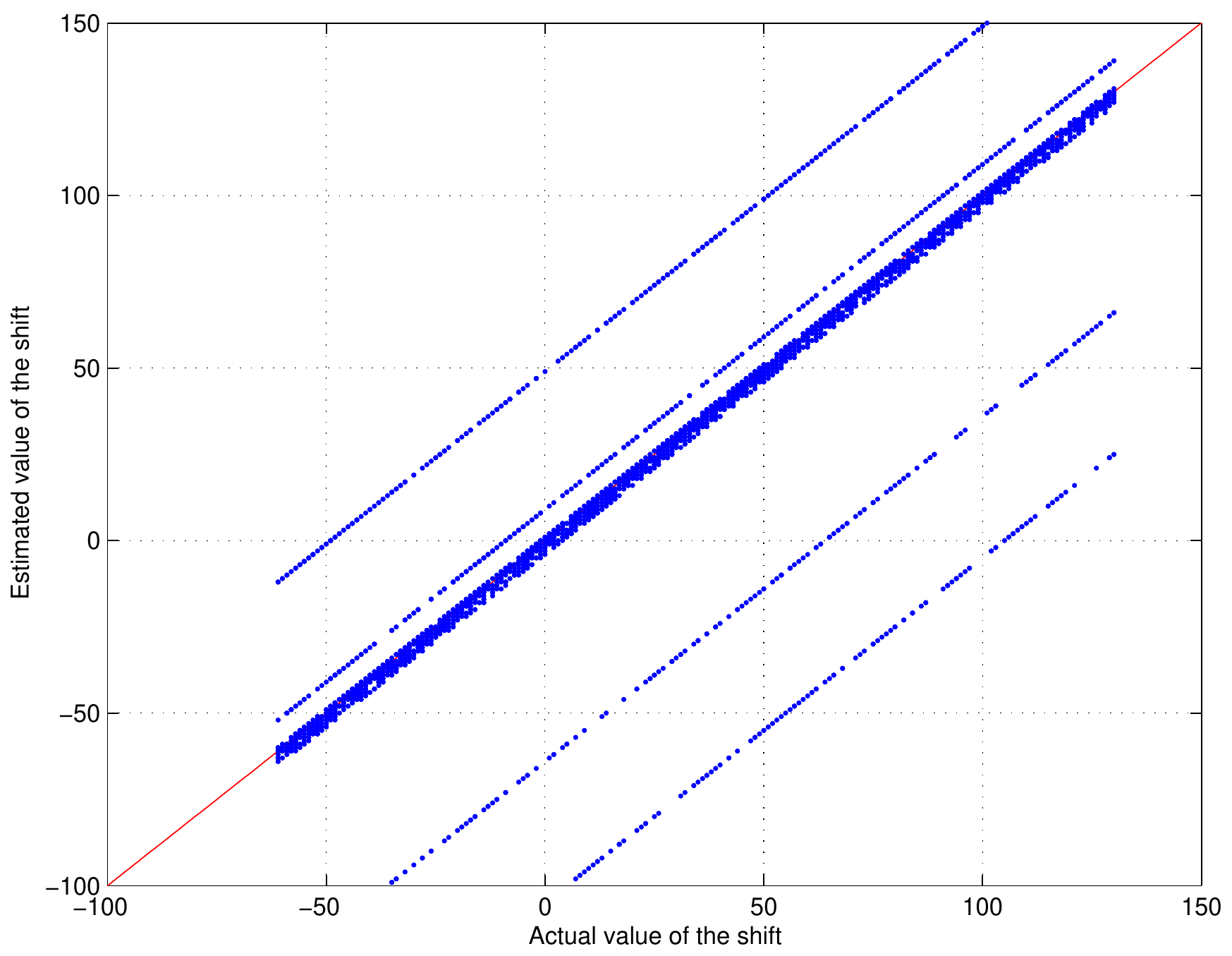}}
\caption{Results for K=200 and $\sigma^2 = 0.1$; (a) two curves before alignment. (b) comparison between estimated against actual values (blue dots) of the shifts for $\lambda = 50$: good estimates must be close to the identity line (red curve). (c) comparison between estimated and actual values of the shifts for $\lambda = 10$.}
\label{fig:results_low_noise}
\end{figure*}
\label{sec:applications}

We present in this section results based on simulations and real data. Since we provide a generic method suitable for most biological signals, we focus in our simulations on a neuroscience model, while our real datasets stem from the ECG framework. In the latter case, we compare our method to the one described in~\cite{silverman:ramsay:2005} which is often used by practitioners, that is a measure of fit based on the squared distance between the average pulse and the shifted pulses leading to a standard Least Square Estimate of the shifts. We present in our simulations results for several values of $K$. However, a method for choosing automatically the parameter $K$ has been suggested in~\cite{trigano:isserles:ritov:2008}: since the term $A_M(k)$ can be built iteratively and converges when the number of curves tends to infinity due to Remark~\ref{rem:convergence_A}, $K$ can be chosen such that
$$
K \eqdef \min \left\{ L \ ; \ \sum_{k \in {\cal K}} \nu_k \left(A_M(k) - \frac{1}{L+1}\sum_{l=0}^L |f_{k,l}|^2 \right)^2 \leq \epsilon \right\} \eqsp ,
$$
where $\epsilon$ is a precision threshold fixed by the user. It is however obvious that the optimal choice of $K$ should depend on the functional properties of the signal $s$, which are unknown in a semiparametric framework.

\subsection{Simulations results}

 Using simulations we can study the influence of the parameters $K$ and $\lambda$ empirically by providing the Mean Integrated Squared Error (MISE) for different values of $K$ and $\sigma^2$. We use a fixed number of blocks $N=20$. The weighting parameter is chosen as $\lambda=[K^\beta]$, where $0<\beta<1$. Choosing $\beta$ close to $1$ enables us to align the curves of a given block with respect to the reference curve.

\subsubsection{Experimental protocol}

Simulated data are created according to the discrete model~\eqref{eq:shifted_model}, and we compute the estimators for different values of the parameters $K$, $\lambda$ and $\sigma^2$. For each curve, we sample  $512$ points equally spaced on the interval $[0;2\pi]$. We make the experiment with $s$ computed according to the standard Hodgkin-Huxley model for a neural response. The shifts are drawn from a uniform distribution $\,\mathcal{U}(120\pi/256 ,325\pi/256)$, and $\theta_0 = \pi$. The sequence $\{\nu_k, \ k \in \mathcal{K}\}$ is taken such that $\nu_k = 1$ for $k=-\frac{150}{2} \ldots  \frac{150}{2}$ and $\nu_k = 0$ otherwise. Though this choice is not optimal, it provides sufficiently good results on the present simulations to illustrate our purpose. Details on the problem of the choice of the tapering sequence $\{\nu_k, \ k \in \mathcal{K}\}$ may be found in~\cite{gamboa:loubes:maza:2005}.

\subsubsection{Results}

\begin{figure}[ht]
\centering
\includegraphics[width=0.8\linewidth]{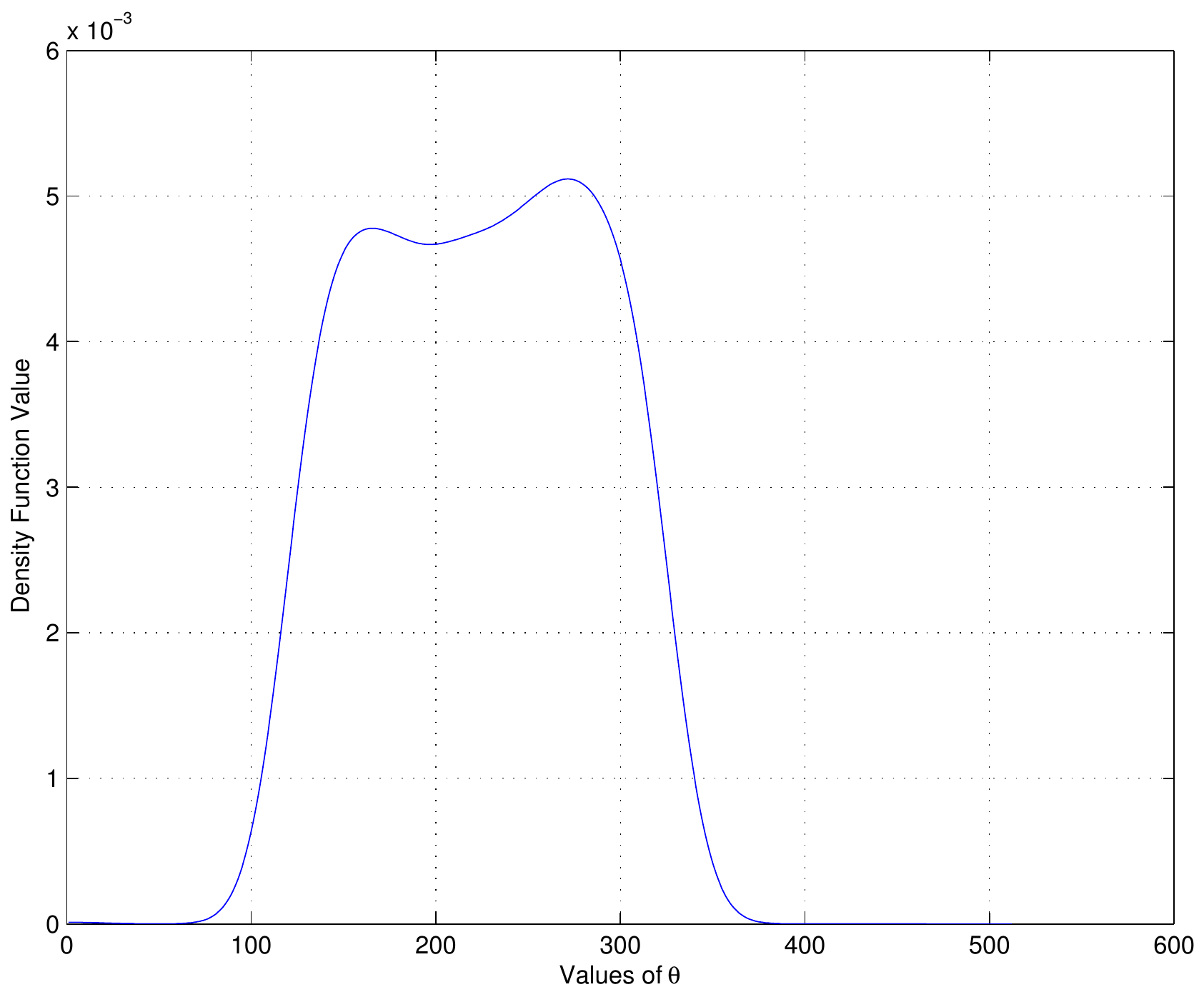}
\caption{Probability density estimation for $N=20$, $K=200$ and $\sigma^2 = 0.1$.}
\label{fig:pdf_estimation}
\end{figure}

\begin{figure}[th]
\centering
\includegraphics[width=0.8\linewidth]{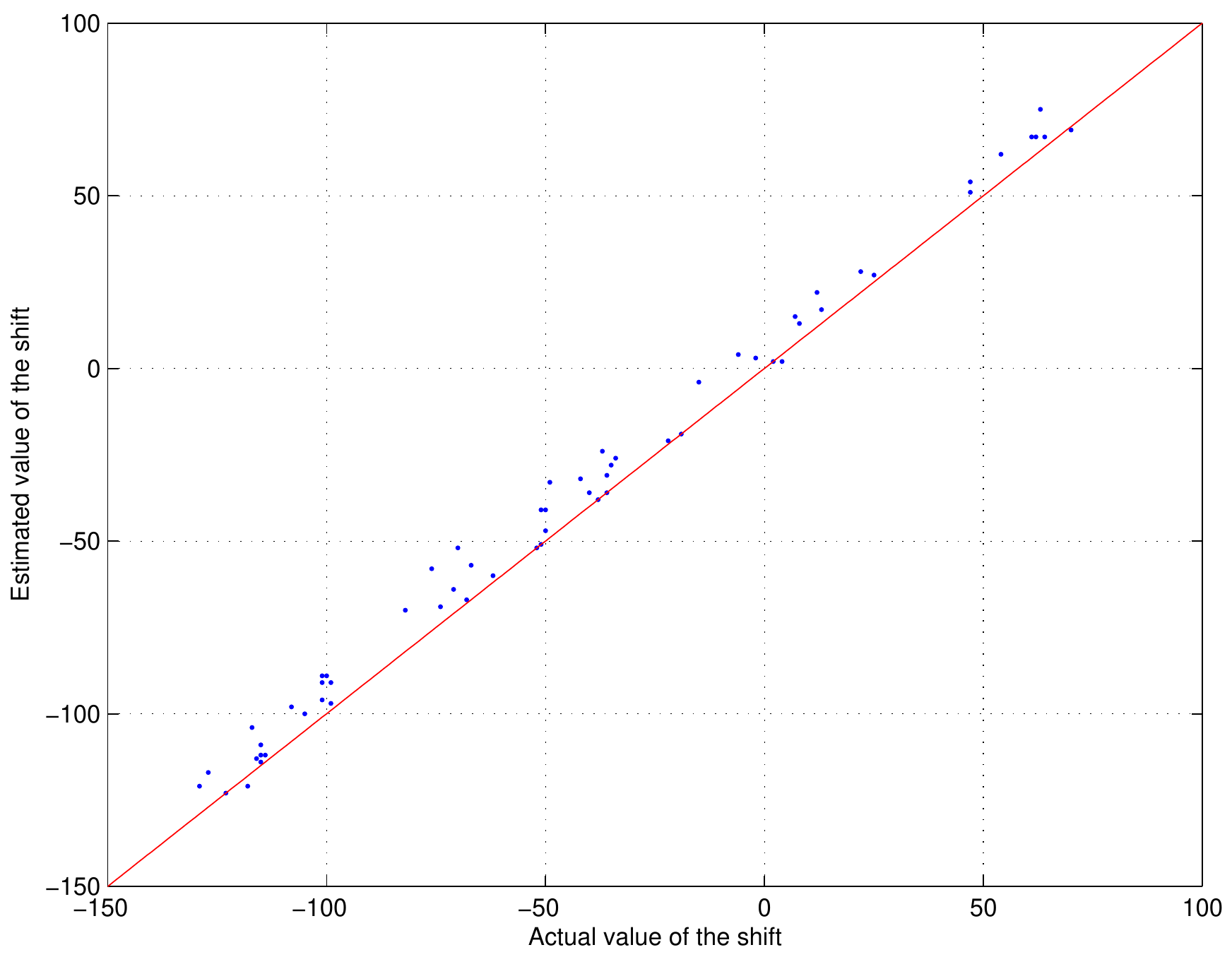}
\caption{Shift estimation using Least Square Estimate (see~\cite{silverman:ramsay:2005}) for one block.}
\label{fig:kneip}
\end{figure}

We present in Figure~\ref{fig:results_low_noise} results obtained using the alignment procedure, in the case of high noise level ($\sigma^2=0.1$). We also compare our estimations with those obtained with an existing method, namely curve alignment according to the
comparison between each curve to the mean curve~\cite{silverman:ramsay:2005}. Results using landmark alignment are
displayed in Figure~\ref{fig:kneip}. We observe that the efficiency of this approach is less than our estimate achieves with \(\lambda=50\), Figure~\ref{fig:results_low_noise}-(b), but is better than the estimate with \(\lambda=10\), Figure~\ref{fig:results_low_noise}-(c). An example of density estimation is displayed in Figure~\ref{fig:pdf_estimation}, using a
Gaussian kernel. It should be noticed, however, that $h$ is a free parameter which may exhibit a strong influence on the resulting estimate. A too small value of $h$ leads to over-fitting, whereas taking large values of $h$ leads to hide the multimodality of $f$, if any. Choosing a data-driven bandwidth selection of $h$ is thus far from trivial, and out of the scope of this paper: we refer to~\cite{berlinet:devroye:1994,jones:marron:1996} for an in-depth description of the existing procedures. The bandwidth $h$ is chosen by Silverman's ``rule-of-thumb''~\cite{silverman:1986}. We retrieve the uniform distribution of
$\theta$. Table~\ref{tab:MISE} shows the estimated MISE for different values of $K$ and $\sigma^2$, with $\lambda=[K^{0.9}]$ and \(N=100\) blocks. The first given number is the value for our estimate, while the second is for the estimator of \cite{silverman:ramsay:2005}.  Note the dominance of the proposed estimator in all cases, in particular for the more noisy situations.

\begin{table}
\centering
\begin{tabular}{|l|c|c|c|c|c|} \hline
$\sigma^2$  & K=10 & K=20 & K=30 & K=50 & K=100 \\ \hline
\multirow{2}{*}{$ 0$}  &{\small  0.0305 }&{\small  0.0228 }&{\small  0.0198 }&{\small  0.0153 }&{\small  0.0106} \\
 &{\small  0.0306 }&{\small  0.0234 }&{\small  0.0199 }&{\small  0.0156 }&{\small  0.0109} \\
\hline
\multirow{2}{*}{$ 10^{-4}$}  & {\small 0.0312} & {\small 0.0218} & {\small 0.0183} & {\small 0.0156} & {\small 0.0121} \\
 & {\small 0.0325} & {\small 0.0232} & {\small 0.0212} & {\small 0.0183} & {\small 0.0158} \\
\hline
\multirow{2}{*}{$ 10^{-2}$}  & {\small 0.0296} & {\small 0.0218} & {\small 0.0172} & {\small 0.0143} & {\small 0.0120} \\
 & {\small 0.0306} & {\small 0.0232} & {\small 0.0192} & {\small 0.0172} & {\small 0.0143} \\ \hline
\multirow{2}{*}{$ 1$}  & {\small 0.0326} & {\small 0.0274} & {\small 0.0248} &  {\small 0.0255} &  {\small 0.0288}\\
 & {\small 0.0547} & {\small 0.0806} & {\small 0.0514} & {\small 0.0553} & {\small 0.0741} \\
\hline
\end{tabular}
\caption{The MISE of the two density estimates.}
\label{tab:MISE}
\end{table}

\subsection{Results on real data}

We now compare the estimated average aligned signal of the two methods applied to ECG signals. The data was obtained from the Hadassah Ein-Karem hospital.

\subsubsection{Experimental protocol}

In order to obtain a series of heart cycles, we first make a preliminary segmentation using the method of~\cite{gasser:kneip:1995}, namely alignment according to the local maxima of the heart cycle. We then apply our method, and compare it to the alignment obtained by comparing the mean curve to a shifted curve one at a time. We took in this example $K=30$ and $\lambda=K^{0.75}$.

\subsubsection{Results}

\begin{figure*}[ht]
     \centering
     \subfigure[Aligned heart cycles and average signal (black dotted
     curve) using the standard method]{\label{fig:fda_align} \includegraphics[width=.4\textwidth]{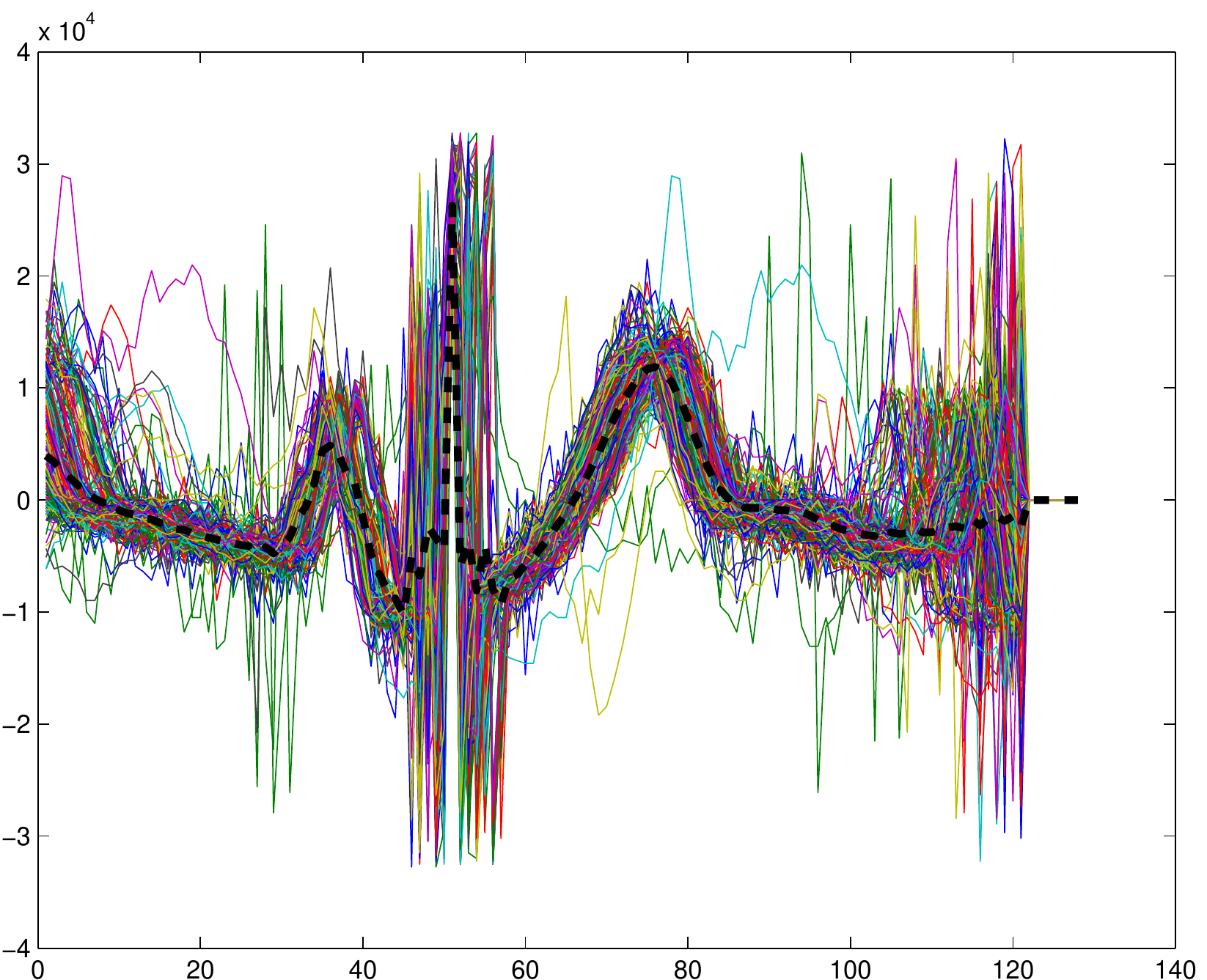}}
     \hspace{.3in}
     \subfigure[Aligned heart cycles and average signal (black dotted
     curve) using the proposed method]{
          \label{fig:semip_align} \includegraphics[width=.4\textwidth]{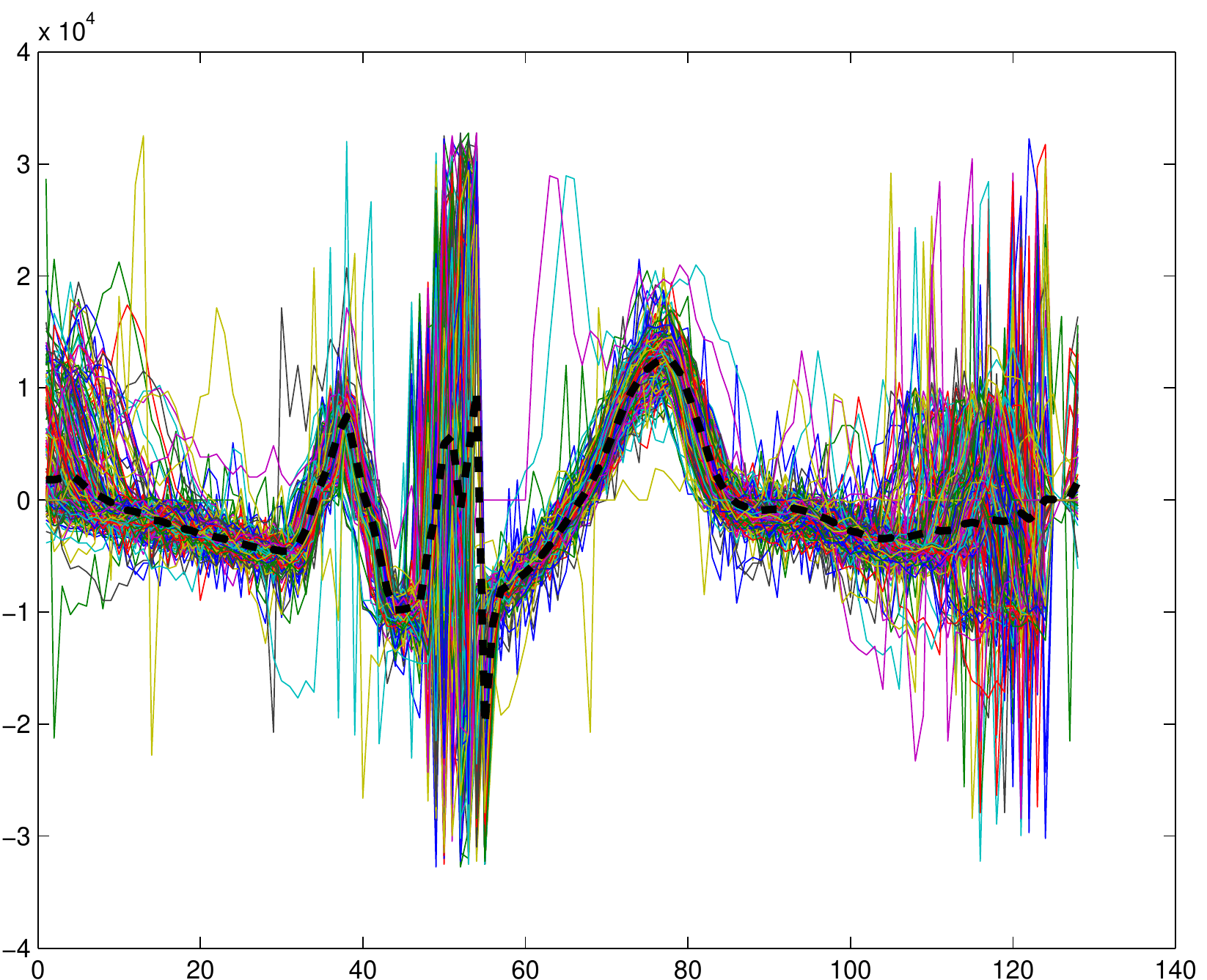}}\\
     %\vspace{.3in}
     \subfigure[Aligned heart cycles using the standard method, zoom
     for the first 30 curves]{\label{fig:FDA_zoom}
       \includegraphics[width=.4\textwidth]{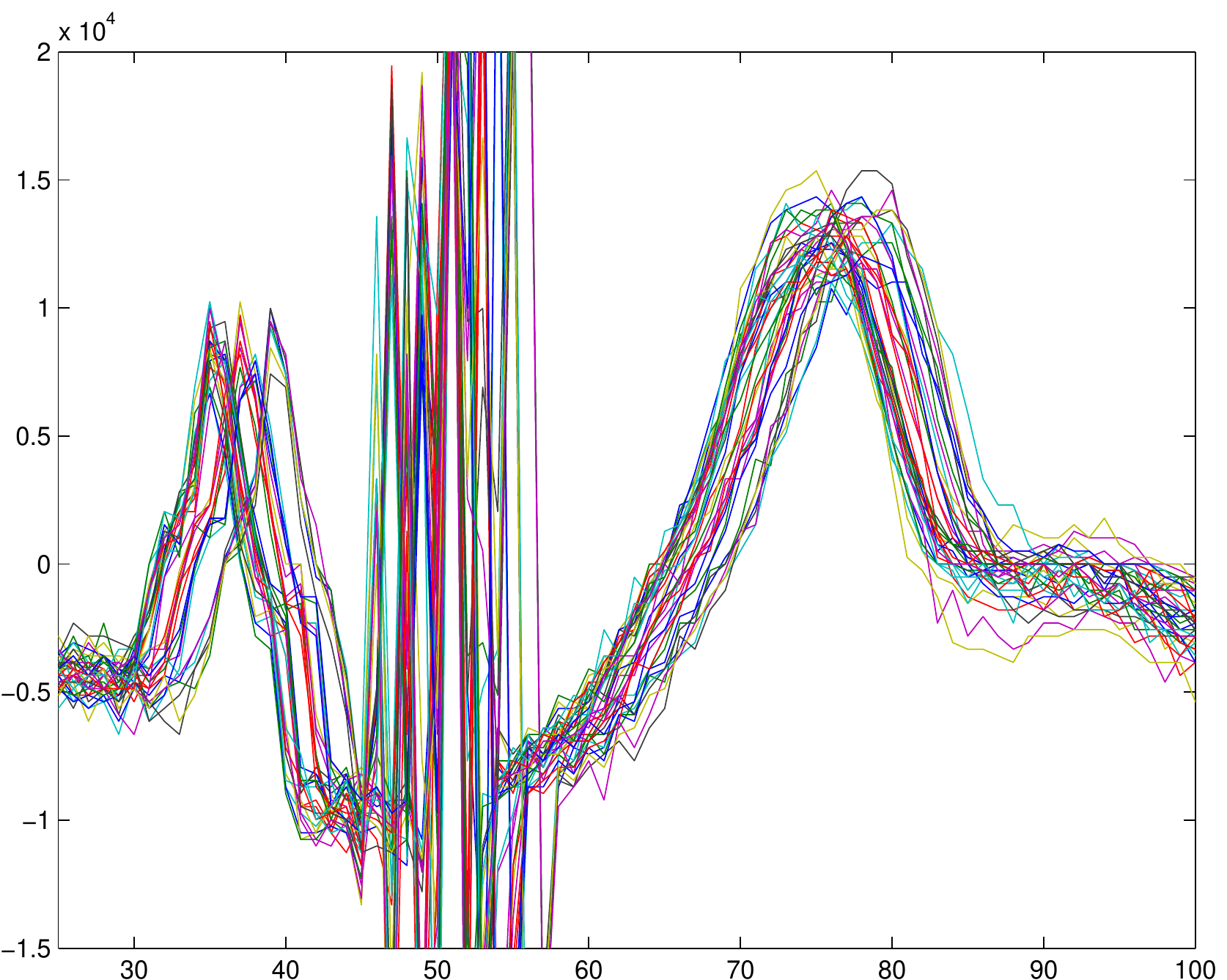}}
     \hspace{.1in}
     \subfigure[Aligned heart cycles using the proposed method, zoom for the first 30 curves]{
           \label{fig:semip_zoom} \includegraphics[width=.4\textwidth]{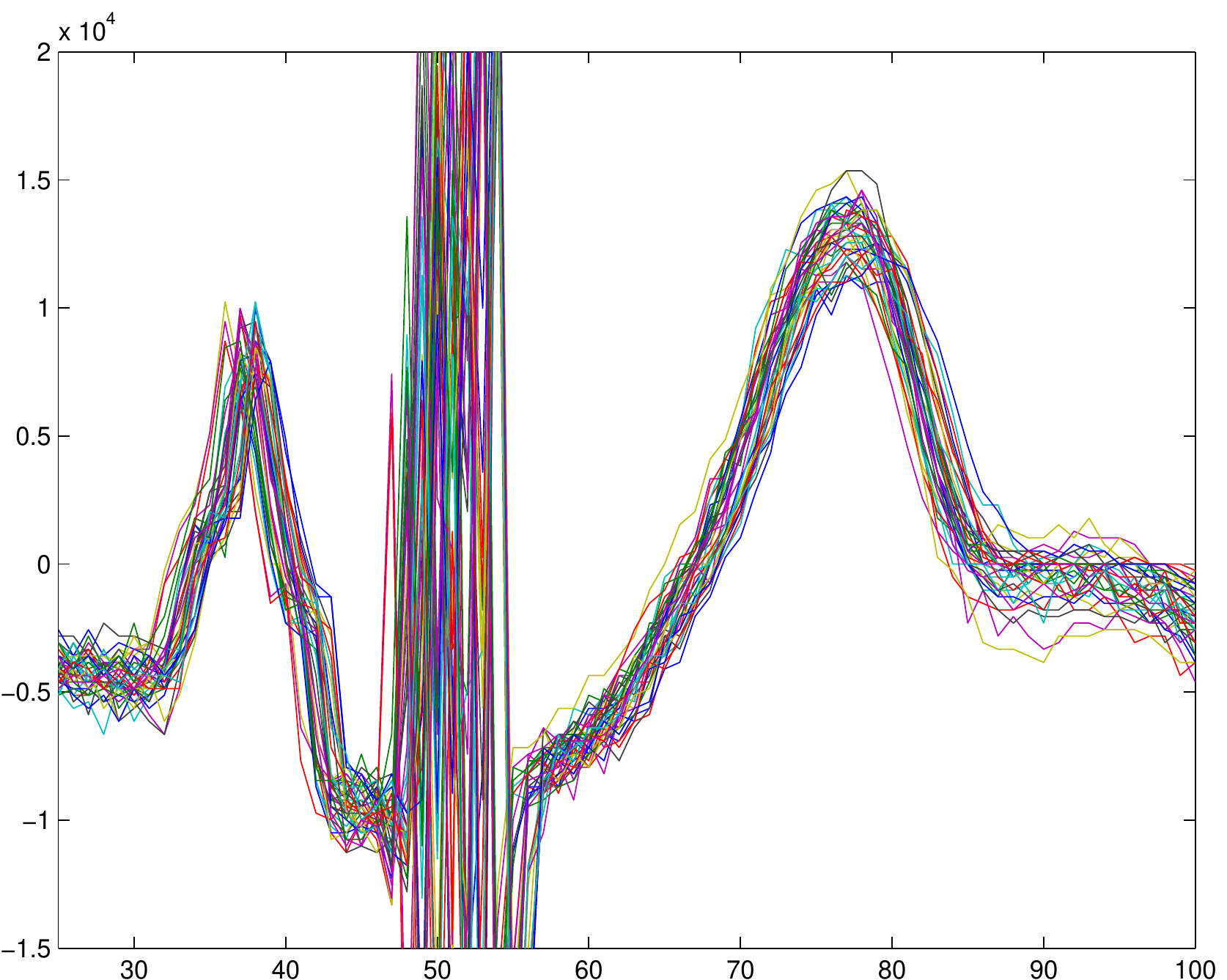}}
     \caption{Comparison between the state-of-the-art and the proposed
       method for the alignment of heart cycles (arbitrary units). A
       semiparametric approach appears more appealing to align cycles
       according to their starting point, and allows to separate more
       efficiently to P-wave, the QRS complex and the T-wave.}
     \label{fig:results_real}
\end{figure*}

The results  are presented in Figure~\ref{fig:results_real}. Comparison of Figures~\ref{fig:FDA_zoom} and~\ref{fig:semip_zoom} shows that the proposed method outperforms the standard one. Moreover, when computing the average of the reshifted heart cycle, we observe that our method allows to separate more efficiently the different parts of the heart cycle; indeed, the separation between the P-wave, the QRS-complex and the T-wave are much more visible, as it can be seen by comparing the average signals obtained in Figure~\ref{fig:fda_align} and Figure~\ref{fig:semip_align}.

\subsection{Influence of ECG perturbations on the proposed algorithm}

As we saw, the model fits reasonably well the data we have at hand, and in fact perform better than the competing algorithm. The ideal model may not fit other data sets in which the shape of the heart pulse changes, or additional perturbations occur. Although no estimation procedure can operate under any possible distortion of the data, we now show that our procedure is quite robust against the main type of potential distortions. The main type of perturbations related to the processing of ECG data  are of four kinds (cf. \cite{sayadi:shamsollahi:2007}):

\begin{itemize}
\item the baseline wandering effect, which can be modeled by the addition of a very low-frequency curve.
\item 50 or 60 Hz power-line interference, corresponding to the addition of an amplitude and frequency varying sinusoid.
\item Electromyogram (EMG), which is an electric signal caused by the muscle motion during effort test.
\item Motion artifact, which comes from the variation of electrode-skin contact impedance produced by electrode movement during effort test.
\end{itemize}

\begin{figure*}[ht]
\begin{tabular}{cc}
\includegraphics[width=0.45\linewidth]{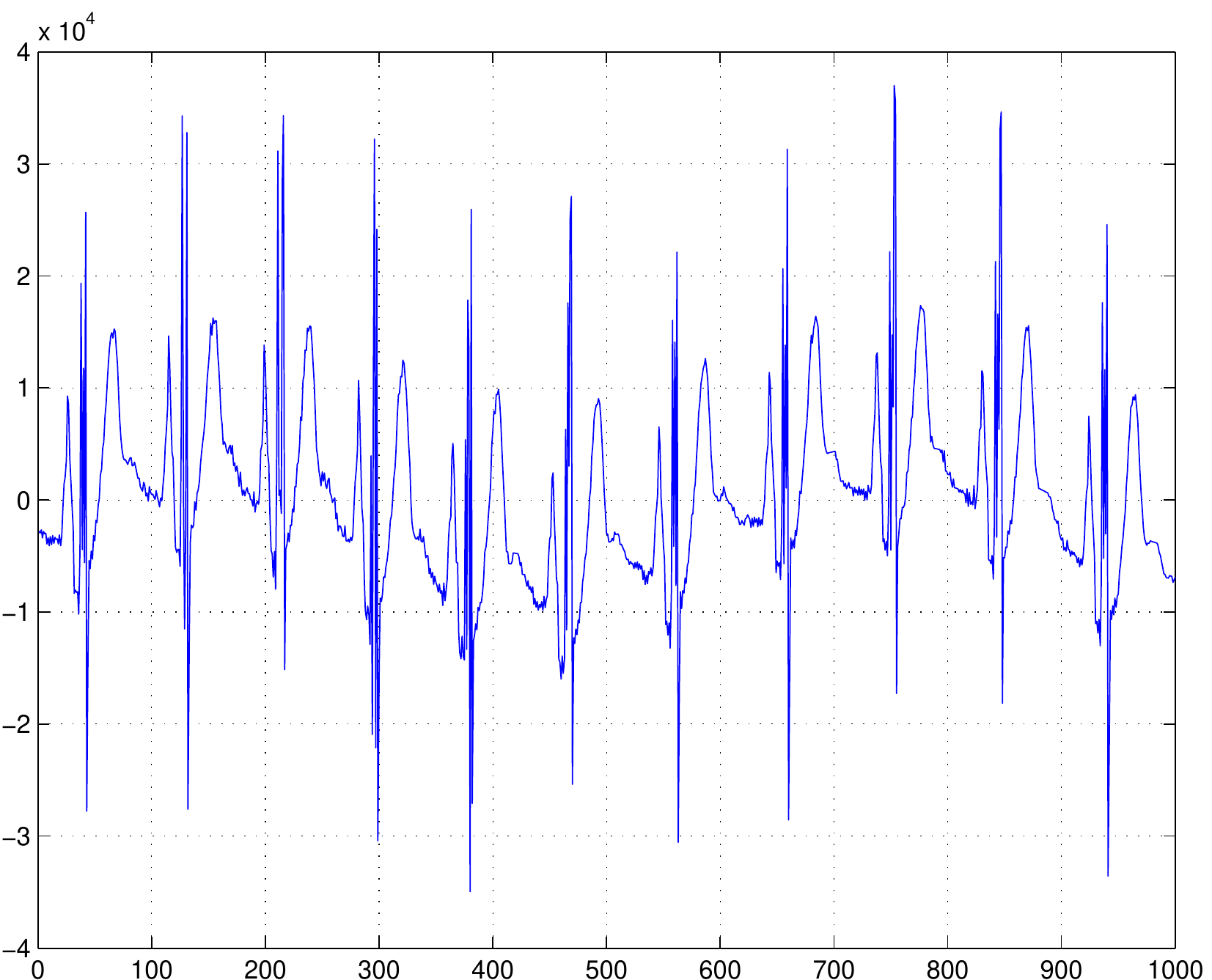}&\includegraphics[width=0.45\linewidth]{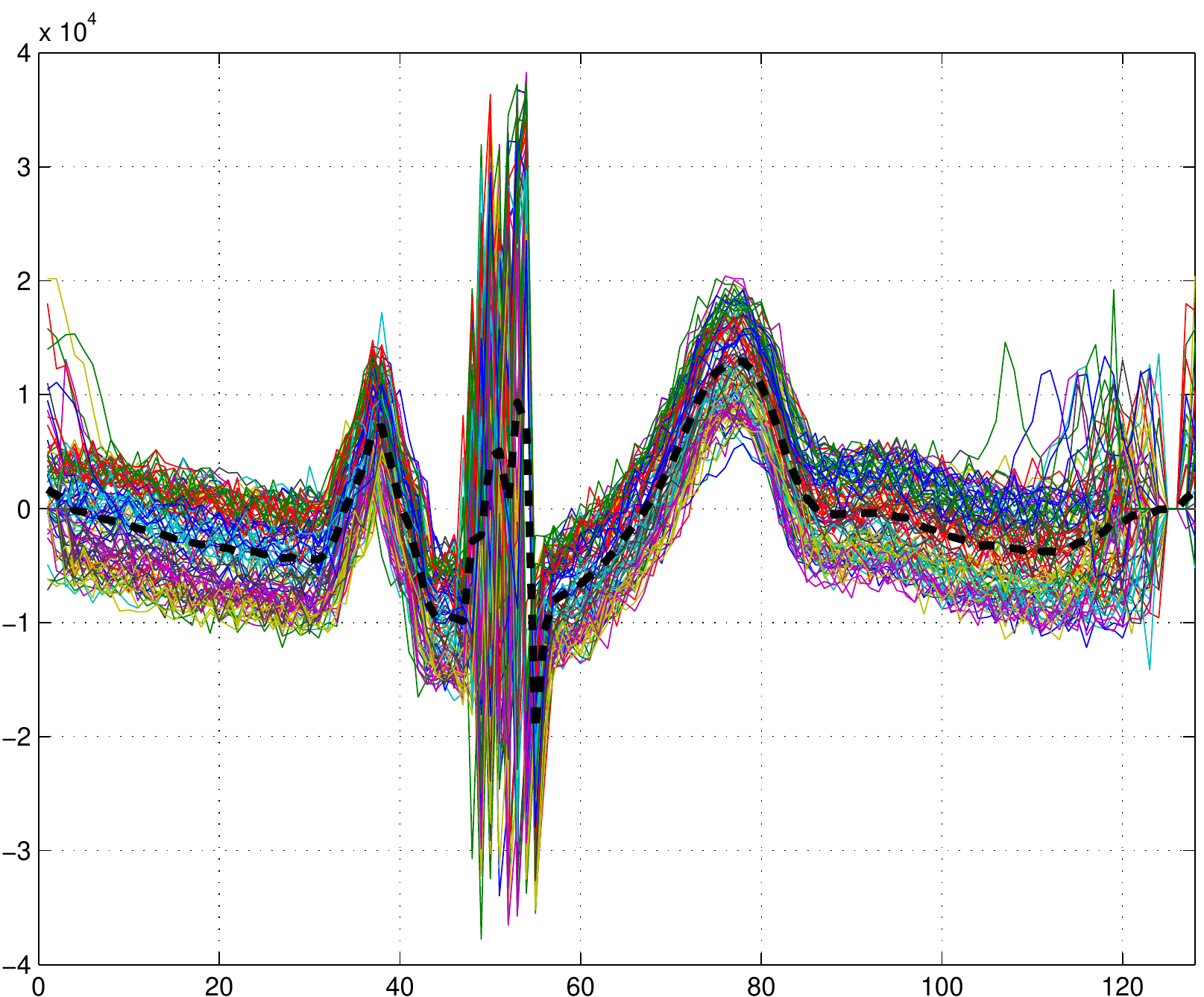}\\
(a)& (b)
\end{tabular}
\caption{Effect of the baseline wander phenomenon over the proposed curve alignment method: distorted signal (a), and aligned pulses with the average ECG pulse obtained for one block (b)}\label{fig:baseline_wander}
\end{figure*}

To keep the discussion within the scope of the paper, we chose to focus on two perturbations, namely the baseline wander effect and the power-line interference effect. We present in Figure~\ref{fig:baseline_wander} the effect of baseline wander on the proposed algorithm. This effect was simulated by the addition of a low-frequency sine to the ECG measurements. We took here $N=100, K=100, \lambda = K^{0.9}$.

We observe that the proposed curve alignment algorithm is robust regarding this kind of perturbations, since we observe well-aligning curves and very little change on the average pulse shape compared to the one obtained without this perturbation. This can be interpreted as follows: since the baseline is in this situation a zero-mean process, the averaging which is done while computing the cost function naturally tends to cancel the baseline. However, we  remark that the baseline wander phenomenon can cripple the preliminary segmentation, if the amplitude of the baseline is too high. This problem can be easily circumvented by means of a baseline reduction prefiltering, such as proposed in~\cite{sayadi:shamsollahi:2007,mneimneh:yaz:2006,mozaffary:tinati:2005}.

\begin{figure*}[ht]
\begin{tabular}{cc}
\includegraphics[width=0.45\linewidth]{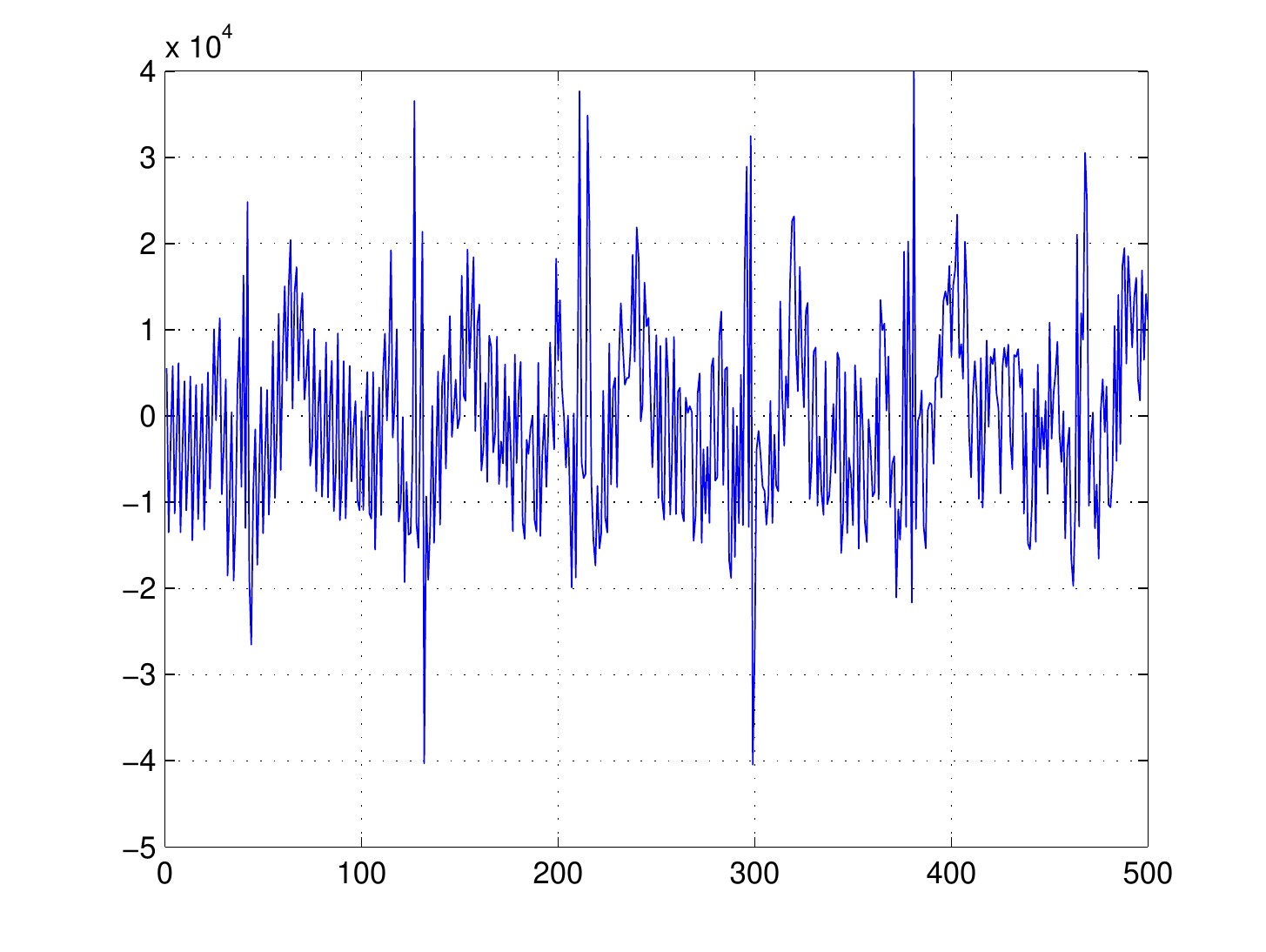}&\includegraphics[width=0.45\linewidth]{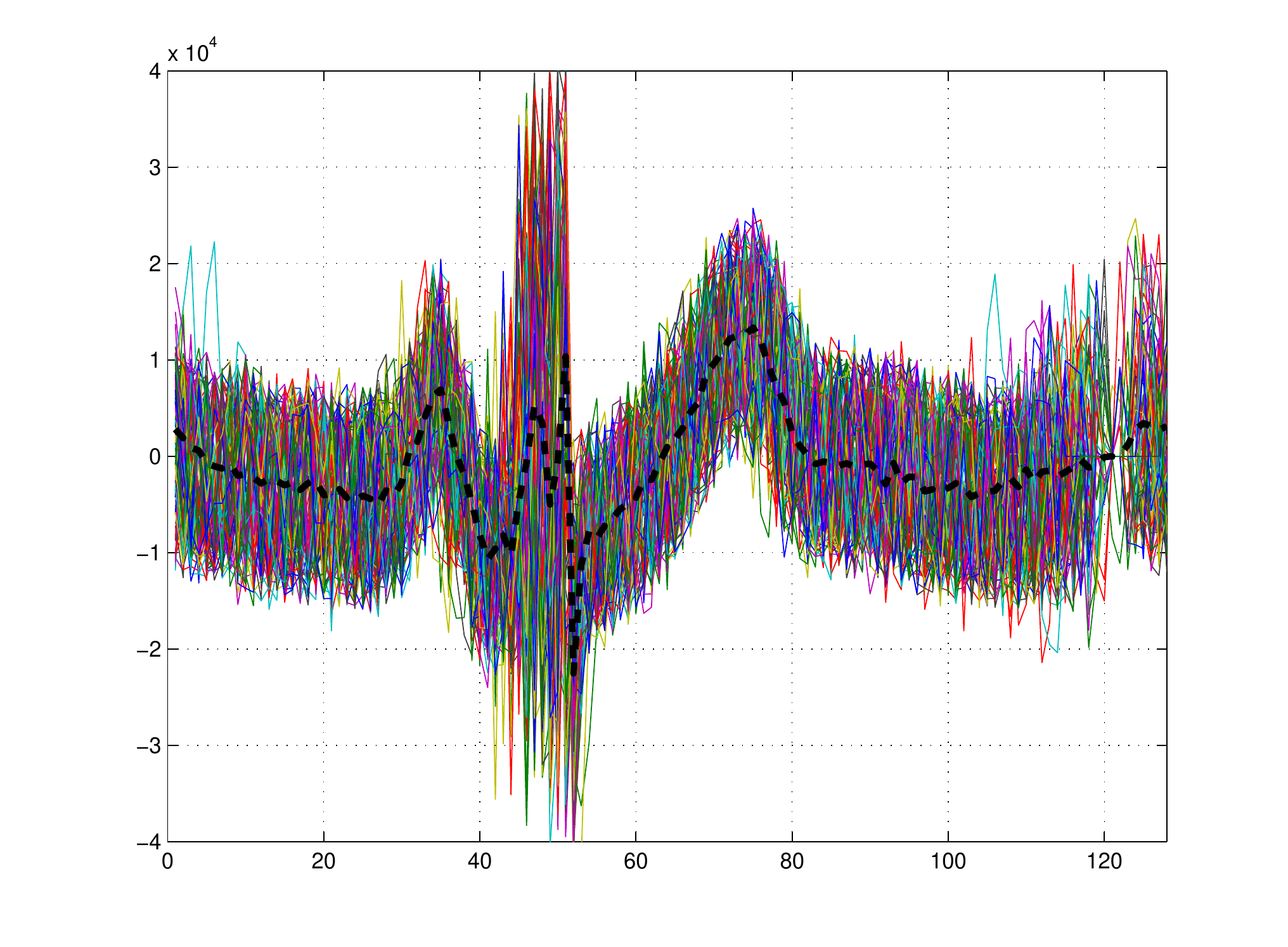}\\
(a)& (b)
\end{tabular}
\caption{Effect of the power-line interference phenomenon over the proposed curve alignment method: distorted signal (a), and aligned pulses with the average ECG pulse obtained for one block (b)}
\label{fig:power-line}
\end{figure*}

We now consider the problem of power-line interference. In order to artificially simulate the original signal with a simulation of the power-line interference, we used the model described in~\cite{avendano:avendano:2007}, that is we add to the ECG signal the following discrete perturbation:
\begin{equation*}
y[n]=(A_0+\xi_A[n])\sin\left(\frac{2\pi(f_0+\xi_f[n])}{f_s}n\right)\ ,
\end{equation*}
where $A_0$ is the average amplitude of the interference, $f_0$ its frequency, $f_s$ the sampling frequency of the signal and $\xi_A[n],\xi_f[n]$ are white Gaussian processes used to illustrate possible changes of the amplitude and frequency of the interference. The results of the curve alignment procedure are presented in Figure~\ref{fig:power-line}, for a similar choice of $N,K$ and $\lambda$.

\begin{figure}
\centering
\includegraphics[width=0.8\linewidth]{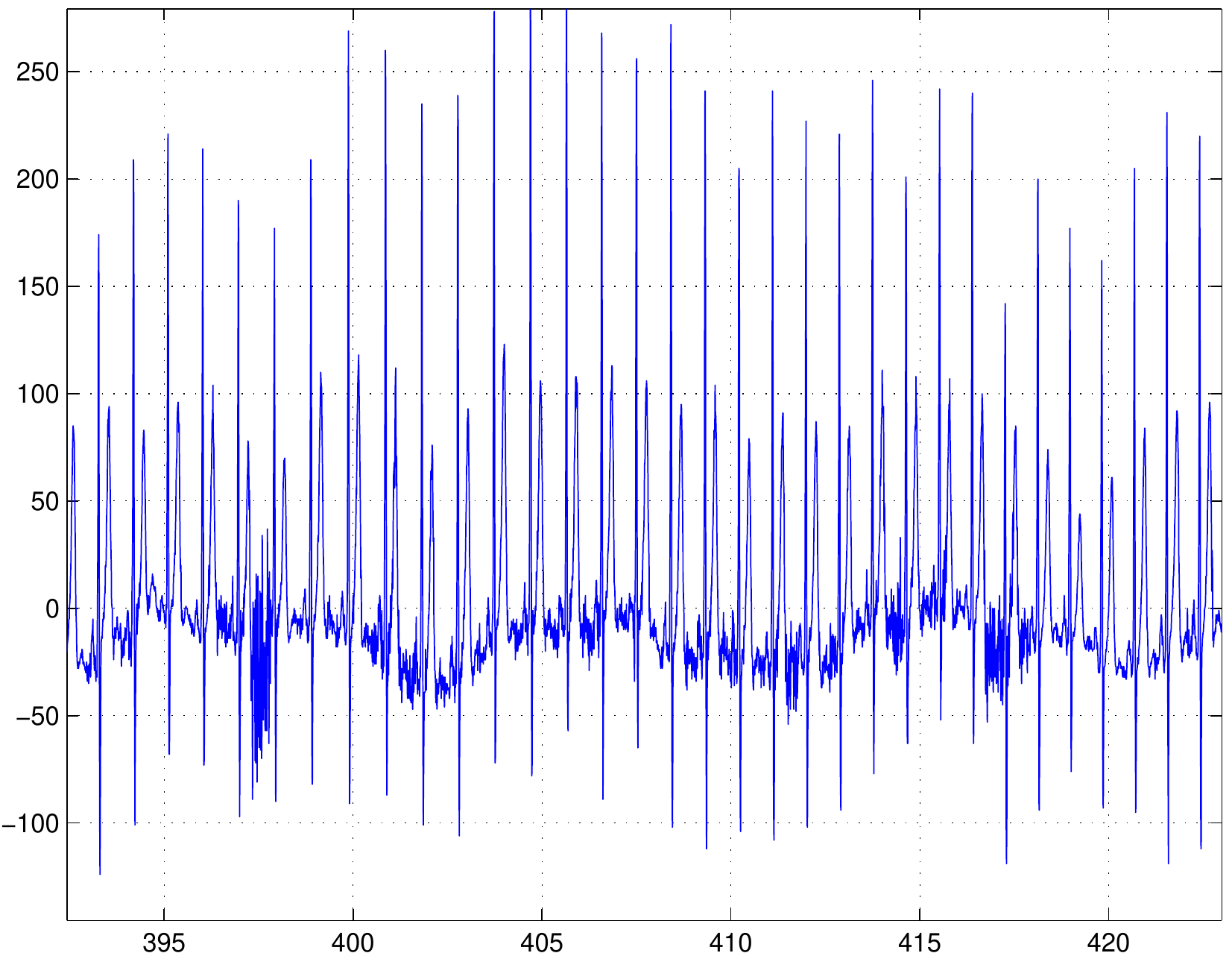}
\caption{ECG signal with real baseline wander and power-line interference (partial)}
\label{fig:real_pwl_base}
\end{figure}

\begin{figure}
\centering
\includegraphics[width=0.8\linewidth]{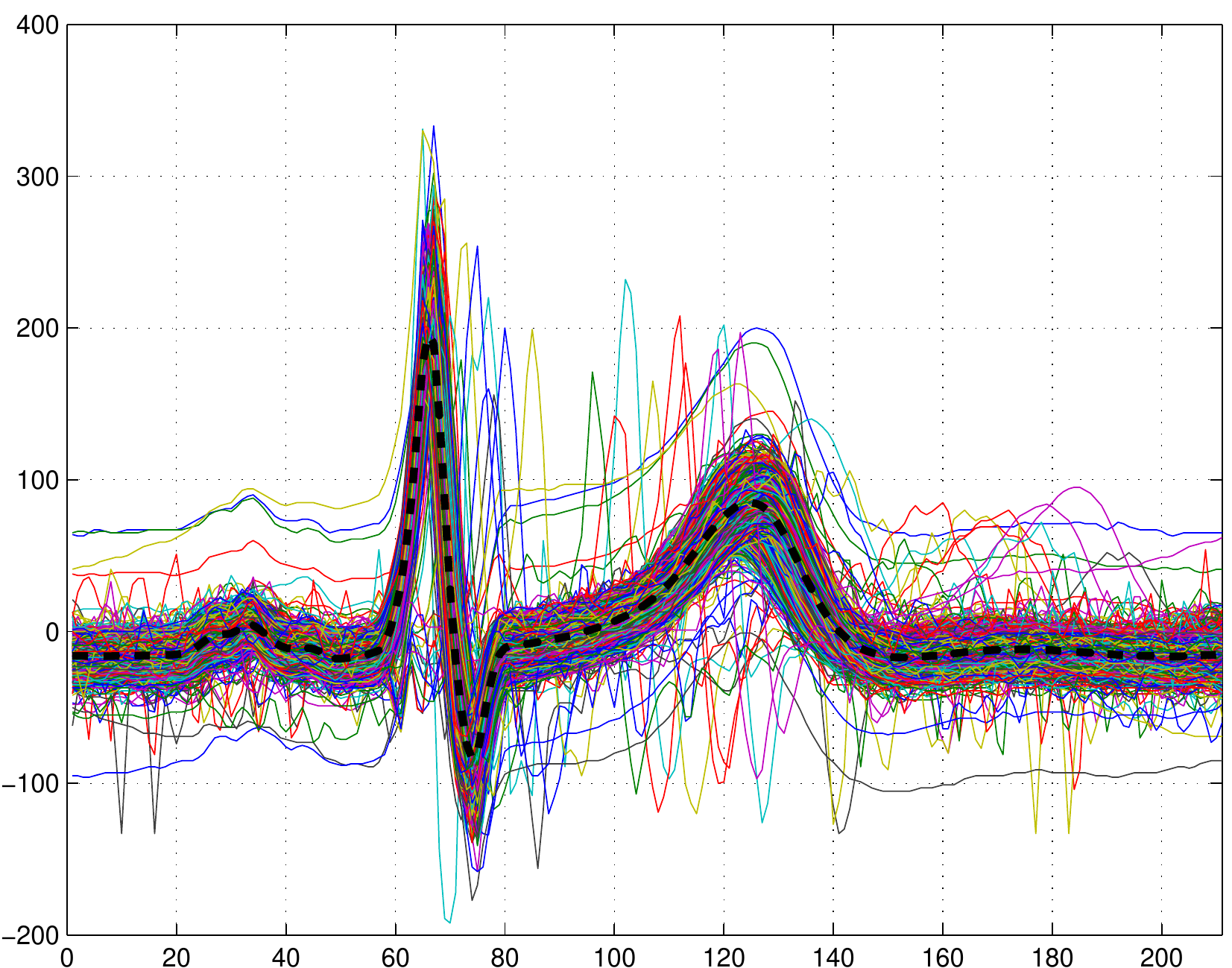}
\caption{Obtained curves before the curve alignment procedure and associated average signal (dotted).}
\label{fig:pre_algo}
\end{figure}

\begin{figure}
\centering
\includegraphics[width=0.8\linewidth]{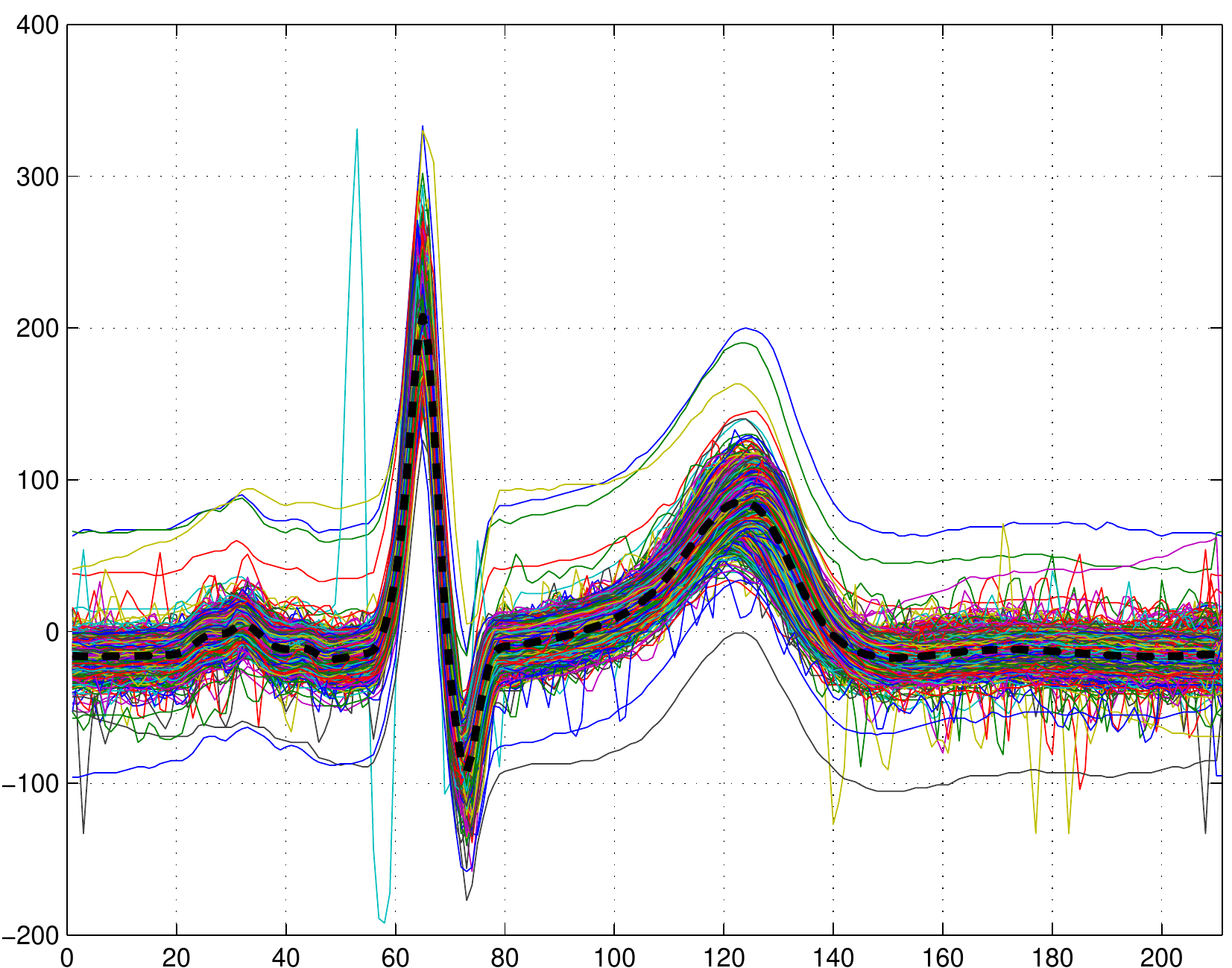}
\caption{Aligned curves by means of the proposed method, and average curve (dotted).}
\label{fig:after_algo}
\end{figure}

As shown in Figure~\ref{fig:power-line}, the proposed algorithm is robust for this kind of distortion, as we retrieve about the same average signal after alignment of the curves. It shall be noted, once again, that this kind of perturbation can interfere with the segmentation procedure, and that for interferences with high amplitude, a prefiltering step as described in~\cite{christov:2000,levkov:mihov:2005,ziarani:konrad:2002} could be applied. Both results illustrate the robustness of semiparametric methods for curve alignment, when compared to standard FDA analysis.
We now apply the proposed algorithm to a real ECG signal displayed in Figure~\ref{fig:real_pwl_base}, which is distorted by power-line interference and baseline wander.
After a preliminary segmentation, we get the individual pulses displayed in Figure~\ref{fig:pre_algo}. The aligned curves and the obtained average signal are presented in Figure~\ref{fig:after_algo}. It can be noted that the proposed method still performs well and is robust to aformentioned perturbations. The average signal obtained by the proposed algorithm is therefore more representative.

\subsection{Discussion}

Figures~\ref{fig:results_low_noise}(b) and \ref{fig:results_low_noise}(c) are a good illustration of Proposition~\ref{prop:bad_control_deterministic}. Figure~\ref{fig:results_low_noise}(c) shows that when \(\lambda\) is too small, the curves are well aligned within the blocks, but blocks have different constant shifts. Taking a larger $\lambda$  addresses this problem, as it can be seen in Figure~\ref{fig:results_low_noise}(b). Our proposed method uses all the available information and not only the information contained in the neighborhood of the landmarks. The advantage of our method is evident with noisy curves, when locating the maximum of each curve is very difficult.

Not surprisingly, the number of curves in each block $K$ may be low if the noise variance remains very small (first column of  Table~\ref{tab:MISE}), the limiting case $K=2$ consisting in aligning the curves individually. Theoretically,   \(K\) should be taken as large as possible. However, this comes with a price, the largest the \(K\) the more difficult is the optimization problem.

% M-estimation for curve alignment is also discussed  in~\cite{gamboa:loubes:maza:2005}. In fact, \cite[Theorem 2.1]{gamboa:loubes:maza:2005}
% shows that a statistically consistent alignment can be obtained only
% when filtering the curves and aligning the low-frequency
% information. Therefore, an approach based on the spectral information
% is more likely to achieve good alignment by comparison to the standard method of~\cite{silverman:ramsay:2005}.  Still, the choice of the parameter $K$ of our method is easier than the choice of the sequence $\{\delta_j, \ j\in\zset\}$ needed for the estimator described in~\cite{gamboa:loubes:maza:2005}.

\section{Conclusion}

We proposed in this paper a method for curve alignment and density estimation of the shifts, based on an M-estimation procedure on a functional of the energy spectrum density. The proposed estimator, deduced from blocks of signals of size $K$, showed good performances in simulations, even when the noise variance is high. On real ECG data, the proposed method outperforms the functional data analysis method, thus leading to a more meaningful average signal, which is of interest for the study of some cardiac arrhythmia. Investigations of the associated kernel estimates, with emphasis on rates of convergence, should appear in a future contribution.

\section{Acknowledgments}

We would like the thank anonymous reviewers, whose insightful comments and thorough review greatly contributed to improve the quality and readability of the paper. We are grateful to the French International Volunteer Exchange Program, who partially funded the present work. We would like to thank Y. Isserles for helping comments while writing the paper. This work was funded by an internal research grant of the SCE.

\appendices

\subsection{Computation of the noise-free part}
\label{sec:noise-free-computation}

If the curves are perfectly aligned, that is if
$\boldsymbol{\alpha}_1=\boldsymbol{\theta}_1$, equation~\eqref{eq:computation_Bjalpha} becomes
\begin{align}%\begin {equation} \begin{split}
&B_1(k,\boldsymbol{\theta}_1)=\frac{|c_s(k)|^2}{(\lambda+K)^2}\sum_{l,m=0}^K \lambda_l
\lambda_m
\nonumber\\
& + \frac{\sigma^2}{n(\lambda+K)^2}\sum_{l,m=0}^K \lambda_l
 \lambda_m \{\rme^{\rmi k (\theta_l-\theta_m)}\times
 \label{eq:computation_Bjtheta}\\
&[V_{k,l}V_{k,m}
 + W_{k,l}W_{k,m} +\rmi (V_{k,l}W_{k,m}
 - W_{k,l}V_{k,m}) ]\}
 \nonumber\\
&+\frac{\sigma c_s(k)}{\sqrt{n}(\lambda+K)^2}\sum_{l,m=0}^K
\lambda_l \lambda_m \rme^{-\rmi k\theta_m}(V_{k,m}-\rmi W_{k,m})
\nonumber \\
& +\frac{\sigma c_s^*(k)}{\sqrt{n}(\lambda+K)^2}\sum_{l,m=0}^K
\lambda_l \lambda_m \rme^{\rmi k \theta_l} (V_{k,l}+\rmi W_{k,l})
\nonumber
\end{align}%\end{split}\end{equation}
Equation~\eqref{eq:computation_Bjalpha} can also be expanded, in
order to find an equation close to~\eqref{eq:computation_A}. We find
after some calculations that
\begin{align}%\begin{equation}\begin{split}
& B_1(k,\boldsymbol{\theta}_1)
\nonumber\\
&= |c_s(k)|^2 + \frac{\sigma^2}{n(\lambda+K)^2} \sum_{l=0}^K \lambda_l^2 (V_{k,l}^2 +
W_{k,l}^2)
\nonumber\\
&+ \frac{2\lambda \sigma^2}{n(\lambda + K)^2}
\mathrm{Re}\{\sum_{l=1}^K \rme^{\rmi k \theta_l }[V_{k,l}V_{k,0}
  + W_{k,l}W_{k,0}
  \nonumber\\
& \hspace{3em}+ \rmi (V_{k,l}W_{k,0} - W_{k,l} V_{k,0}) ] \}
 \label{eq:Bjtheta_form2}\\
&+ \frac{2 \sigma^2}{n(\lambda + K)^2}
\mathrm{Re}\{\sum_{1\leq l < m \leq K
  } \rme^{\rmi k \theta_l }[V_{k,l}V_{k,m}
  + W_{k,l}W_{k,m}
  \nonumber\\
&\hspace{2em}+ \rmi (V_{k,l}W_{k,m} - W_{k,l} V_{k,m}) ] \}
  \nonumber\\
&+ \frac{2\sigma \mathrm{Re}(c_s(k))}{\sqrt{n}(\lambda + K)}
\sum_{l=0}^K \lambda_l (V_{k,l} \cos (k\theta_l)-W_{k,l} \sin (k\theta_l))   \nonumber\\
&- \frac{2\sigma \mathrm{Im}(c_s(k))}{\sqrt{n}(\lambda + K)}
\sum_{l=0}^K \lambda_l (V_{k,l} \sin (k\theta_l)+W_{k,l} \cos (k\theta_l))
  \nonumber
\end{align}%\end{split}\end{equation}
%\note{tom}{Maybe simplify terms 2 and 3 in (\ref{eq:Bjtheta_form2})}

Collecting equations~\eqref{eq:computation_A}, \eqref{eq:computation_Bjalpha} and \eqref{eq:Bjtheta_form2}, we can check easily that the only  noise-free part comes from the second sum in~\eqref{eq:exact_approx}, and is equal to $D_1(\boldsymbol{\alpha}_1)$.

\subsection{Proof of Proposition~\ref{prop:noisy-part}}
\label{sec:proof-noisy-part}

Using Equations~\eqref{eq:computation_A} and~(\ref{eq:Bjtheta_form2}),
we get that for all $k$ the deterministic part of $A_M(k) -
B_1(k,\boldsymbol\theta_1)$ vanishes, leading to
\begin{align*}
&A_M(k) - B_1(k,\boldsymbol\theta_1) = \frac{\sigma^2}{(M+1) n}\sum_{l=0}^{M}(V_{k,l}^{2}+W_{k,l}^{2})
\\
&- \frac{\sigma^2}{n(\lambda+K)^2} \sum_{l=0}^K \lambda_l^2 (V_{k,l}^2 +
W_{k,l}^2) \\ % \label{eq:chi2_factors} \\
&- \frac{2\lambda \sigma^2}{n(\lambda + K)^2}
\mathrm{Re}\{\sum_{l=1}^K \rme^{\rmi k \theta_l }[V_{k,l}V_{k,0}
  + W_{k,l}W_{k,0}\\
& ~~~~~~~~~~~~~~~~~~~~~~~~~~+ \rmi (V_{k,l}W_{k,0} - W_{k,l} V_{k,0}) ]
\} \\ %\label{eq:crossed_random1} \\
&\hspace{-2ex}- \frac{2 \sigma^2}{n(\lambda + K)^2}
\mathrm{Re}\{\sum_{1\leq l < m \leq K} \rme^{\rmi k (\theta_l - \theta_m) }[V_{k,l}V_{k,m}
  + W_{k,l}W_{k,m}\\
&~~~~~~~~~~~~~~~~~~~~~~~~~~+ \rmi (V_{k,l}W_{k,m} - W_{k,l} V_{k,m}) ]
\}  \\ %\label{eq:crossed_random2} \\
& +\frac{2\sigma
   \mathrm{Re}(c_s(k))}{(M+1)\sqrt{n}}\sum_{l=0}^{M}(V_{k,l}\cos(k\theta_l)-W_{k,l}\sin(k\theta_l)) \\
& +  \frac{2\sigma \mathrm{Im}(c_s(k))}{(M+1)\sqrt{n}}\sum_{l=0}^{M}(V_{k,l}\sin(k\theta_l)+W_{k,l}\cos(k\theta_l)) \\
&- \frac{2\sigma \mathrm{Re}(c_s(k))}{\sqrt{n}(\lambda + K)}
\sum_{l=0}^K \lambda_l (V_{k,l} \cos (k\theta_l)-W_{k,l} \sin (k\theta_l))\\ & + \frac{2\sigma \mathrm{Im}(c_s(k))}{\sqrt{n}(\lambda + K)}
\sum_{l=0}^K \lambda_l (V_{k,l} \sin (k\theta_l) +W_{k,l} \cos (k\theta_l)).
\end{align*}
\def\LL{{\mu_2}}
Observe that the first two terms are a linear combination of independent Gamma distributed random variables, with second moments $\displaystyle \frac{\sigma^4 (2M+2)(2M+4)}{(M+1)^2 n^2}$ and $\displaystyle \frac{4\sigma^4 (\lambda^4 + K + (\lambda^2+K)^2)}{(K+\lambda)^4 n^2}$, respectively.
All the remaining sums in the latter are of i.i.d.\ random variables with mean zero, and all have sub-Gaussian tails.   Consequently, there is a constant \(D\), independent of \(k\), such that when $K\to\infty$, $\lambda\to\infty$
and $\lambda/K \to 0$:
\begin{align*}
&\hspace{-3em}\|A_M(k) - B_1(k,\boldsymbol{\theta}_1)\|_{\mu_2}
\leq  D\left( \frac{\sigma^2}{n} + \frac{\sigma^2}{nK}
+\frac{\sigma |c_s(k)|}{\sqrt{nK}}\right)
\end{align*}
where for any random variable \(X\),  \(\|X\|_{\mu_2}=\sqrt{\E (X^2)}\). Hereafter, \(D\) is the same constant, large enough to keep all the inequalities valid. From the latter inequality, we get that:
$$
\sum_{k\in {\cal K}} \nu_k\bigl(A_M(k)-B_1(k,\boldsymbol{\theta}_1)\bigl)^2
= \mathrm O_\PP \left(\frac{1}{n^2}\right) + \mathrm O_\PP \left(\frac{1}{nK}\right) \eqsp.
$$
We now study the term $R(k,\boldsymbol\alpha_1)$, that is the part of $B_1(k,\boldsymbol{\theta}_1)-B_1(k,\boldsymbol{\alpha}_1)$ which depends on the random variables $V$ and $W$, using their expression in~(\ref{eq:computation_Bjalpha}) and~(\ref{eq:computation_Bjtheta}). We get that $R(k,\boldsymbol\alpha_1) = I+II+III$, where
\begin{equation*}
I\eqdef\frac{\sigma^2}{n(\lambda+K)^2} \left|\sum_{l=0}^K \lambda_l \rme^{\rmi
    k \alpha_l} (V_{k,l} + \rmi W_{k,l})\right|^2
\end{equation*}
\begin{equation*}
II\eqdef - \frac{\sigma^2}{n(\lambda+K)^2} \left|\sum_{l=0}^K
  \lambda_l \rme^{\rmi
    k \theta_l} (V_{k,l} + \rmi W_{k,l})\right|^2
\end{equation*}
and
\begin{multline*}
III\eqdef2\mathrm{Re} \{ \frac{c_s(k)\sigma}{\sqrt{n}(\lambda+K)^2}
   \sum_{ l,m=0}^K \lambda_l \lambda_m  \times \\
  (\rme^{\rmi k (\alpha_l-\theta_l-\alpha_m)} - \rme^{-\rmi k
       \theta_m})(V_{k,m}-\rmi W_{k,m}) \}
\end{multline*}
According to Cauchy-Schwarz inequality for Hermitian products:
\begin{multline*}
I \leq \\ \frac{\sigma^2}{n}\left[  \frac{1}{K+\lambda}\sum_{l=0}^K \lambda_l \rme^{\rmi
    k \alpha_l} \rme^{-\rmi k \alpha_l} \right] \left[\frac{1}{K+\lambda} \sum_{l=0}^K \lambda_l ( V_{k,l}^2 + W_{k,l}^2)\right]
 \end{multline*}
Since the first term in parentheses is equal to $1$ and the second is bounded in probability by Markov's inequality, we get that $I=O_\PP (1/n)$, and we get that $II = O_\PP (1/n)$ using a similar argument. Finally, let us define $U_{k,m} \eqdef \rme^{-ik\theta_n}(V_{k,m}-\rmi W_{k,m})$; we get for any real number $c$ that
\begin{equation*}\begin{split}
 & |III| \\
 &\leq
  \frac{2|c_s(k)|\sigma}{\sqrt{n}(\lambda+K)^2} \times\Bigl|
   \sum_{l=0}^K  \sum_{m=0}^K  \lambda_l\lambda_m \Bigl(\rme^{\rmi k (\alpha_l-\theta_l-c -\alpha_m+\theta_m+c)}-1\Bigr)  U_{k,m} \Bigr|  \\
& +  \frac{2|c_s(k)|\sigma}{\sqrt{n}(\lambda+K)} \Bigl|
   \sum_{m=0}^K \lambda_m U_{k,m} \Bigr| \eqsp .
\end{split}\end{equation*}
Since the random variables $\{U_{k,m}, m=0\ldots K\}$ are \iid\  with mean zero and finite variances, we get that
\begin{equation}
\label{ukm}
    \sum_{m=0}^K\lambda_m U_{k,m} = \mathrm O_\PP(K^{1/2}).
\end{equation}
Consequently, using \eqref{ukm} and Cauchy-Schwarz inequality, we obtain:
\begin{equation*}
\begin{split}
  &  |III|\\
  &\leq  \frac{2|c_s(k)|\sigma}{\sqrt{n}(\lambda+K)^2} \left | \sum_{l=0}^K  \sum_{m=0}^K  \lambda_l\lambda_m \rme^{\rmi k (\alpha_l-\theta_l-c -\alpha_m+\theta_m+c)}U_{k,m}  \right| \\
  & ~~~~~~~~~~~~~~~~~~~~~~~~~~~~~~~~~~~~~~~~~~~ +\mathrm O_\PP \left(\frac1{\sqrt{nK}} \right)
     \\
     &\leq \frac{2|c_s(k)|\sigma}{\sqrt{n}(\lambda+K)}\Bigl|\sum_{m=0}^K \lambda_m \rme^{-\rmi k ( \alpha_m-\theta_m-c)}U_{k,m}\Bigr| +\mathrm O_\PP \left(\frac1{\sqrt{nK}} \right)
         \\
     &\leq \frac{|c_s(k)|\sigma}{\sqrt{n}(\lambda+K)}\Bigl|\sum_{m=0}^K \lambda_m (\rme^{-\rmi k ( \alpha_m-\theta_m-c)}-1)U_{k,m}\Bigr|  \\
  & ~~~~~~~~~~~~~~~~~~~~~~~~~~~~~~~~~~~~~~~~~~~ +\mathrm O_\PP \left(\frac1{\sqrt{nK}} \right)
     \\
     &\leq
     \frac{|c_s(k)|\sigma}{\sqrt{n}} \Bigl(\frac1{K+\lambda} \sum_{m=0}^K \lambda_m|\rme^{-\rmi k ( \alpha_m-\theta_m-c)}-1|^2\Bigr)^{1/2} \\
  & ~~~~~~~~~~~~~ \times
     \Bigl(\frac1{K+\lambda}\sum_{m=0}^K\lambda_m |U_{k,m}|^2\Bigr)^{1/2}
     +\mathrm O_\PP \left(\frac1{\sqrt{nK}} \right)
     \\
     &=  \mathrm O_\PP(1) \frac{|c_s(k)|\sigma}{\sqrt{n}}
     \Bigl(\frac1{K+\lambda}\sum_{m=0}^K\lambda_m(\alpha_m-\theta_m-c)^2\Bigr)^{1/2} \\
  & ~~~~~~~~~~~~~~~~~~~~~~~~~~~~~~~~~~~~~~~~~~~
     +\mathrm O_\PP \left(\frac1{\sqrt{nK}} \right).
\end{split}
\end{equation*}

Recall that  \(\sum_{\nu\in{\cal K}} \nu_k\) is bounded.  Equation \eqref{sumsqdiff} is obtained if all the bounds above  are collected, and  Assumption \ref{hyp:finite_energy} is used. Finally, since
\begin{align*}
 &\biggl|\|C_1(\boldsymbol{\alpha}_1\|_\nu^2 - \sum_{k\in \cal K}\nu_k \Delta(k, \boldsymbol\alpha_1)^2\biggr|
 \\
 &= \biggl|\sum_{k\in\cal K}\nu_k (A_m(k)-B_1(k,\boldsymbol\theta_1) + R(k,\boldsymbol\alpha_1) +\Delta(k,\boldsymbol\alpha_1))^2 \\ &~~~~~~~~~~~~~~~~~~ - \sum_{k\in \cal K}\nu_k \Delta(k, \boldsymbol\alpha_1)^2\biggr|
 \\
 &\leq 2 \sum_{k\in\cal K}\nu_k (A_m(k)-B_1(k,\boldsymbol\theta_1) )^2 +2\sum_{k\in\cal K}\nu_k  R(k,\boldsymbol\alpha_1)^2
\\
&+2 \biggl|\sum_{k\in\cal K}\nu_k (A_m(k)-B_1(k,\boldsymbol\theta_1) + R(k,\boldsymbol\alpha_1) ) \Delta(k,\boldsymbol\alpha_1)\biggr|
\end{align*}
For any number $c$, a Taylor expansion up to the second order yields:
\begin{align*}
&\Delta (k,\boldsymbol\alpha_1) = |c_s (k)|^2 \left | \left| \frac{1}{K+\lambda} \sum_{m=0}^K\lambda_m \rme^{\rmi k (\alpha_m-\theta_m - c)} \right|^2 -1  \right | \\
%& \leq 2  |c_s (k)|^2  \left | \left| \frac{1}{K+\lambda} \sum_{m=0}^K \lambda_m \rme^{\rmi k (\alpha_m-\theta_m - c)} \right| -1  \right | \\
& \leq 2  |c_s (k)|^2  \left | \frac{1}{K+\lambda} \sum_{m=0}^K \lambda_m \rme^{\rmi k (\alpha_m-\theta_m - c)} -1  \right | \\
& \leq D k^2 |c_s(k)|^2 \biggl| \frac{1}{K+\lambda} \sum_{m=0}^K \bigl( \lambda_m \rmi  (\alpha_m-\theta_m - c) \\
&~~~~~~~~~~~~~- \frac12 \lambda_m(\alpha_m-\theta_m -c)^2  \rme^{\rmi k(\alpha_m-\theta_m-c)\xi} \bigr) \biggr|\\
&\leq Dk^2 |c_s(k)|^2 \biggl[ \left| \frac{1}{K+\lambda} \sum_{m=0}^K \lambda_m  (\alpha_m-\theta_m - c) \right| \\
&~~~~~~~~~~~~~ + \left|  \frac12 \frac{1}{K+\lambda} \sum_{m=0}^K \lambda_m(\alpha_m-\theta_m -c)^2 \right| \biggr]
\end{align*}
Taking $c=\frac{1}{K+\lambda} \sum_{m=0}^K \lambda_m (\alpha_m-\theta_m)$ will cancel the RHS first order term and give the minimum of the RHS second order term; since $\sum_{k\in\cal K} \nu_k k^2 <\infty$ we get eventually that
$$
\sum_{k\in \cal K}\nu_k \Delta(k, \boldsymbol\alpha_1)
\leq D \inf_c\frac1{K+\lambda}\sum_{m=0}^K\lambda_m(\alpha_m-\theta_m-c)^2\, ,
$$
thus we get the last term of~\eqref{sumsqdiff}.
%
%Note then that  
%\begin{align*}
%&\sum_{k\in \cal K}\nu_k \Delta(k, \boldsymbol\alpha_1)^2
%\\
%&\leq D \inf_c\frac1{K+\lambda}\sum_{m=0}^K\lambda_m(\alpha_m-\theta_m-c)^2
%\end{align*}
% and we obtain the last equality of~\eqref{sumsqdiff} by means of Cauchy-Schwarz inequality.

\subsection{Proof of Proposition~\ref{prop:bad_control_deterministic}}
\label{sec:proof_bad_control_deterministic}

Observe that there exists $\gamma_0$ in $(0,1)$ such that, for all
$x$ in  $[-\pi,\pi]$, we have $\cos x \leq 1-\gamma_0 x^2$. Since we have
\begin{equation*}
\left| \frac{1}{(K+\lambda)} \sum_{0\leq m \leq K} \lambda_l \exp\left(\rmi k \, (\theta_m-\alpha_m) \right)\right| \leq 1 \eqsp,
\end{equation*}
then there exists, according to the assumption, two constants $K_0 \geq 0$ and $c$ such
that, for $K\geq K_0$ and every $k$, we have
\begin{multline}\label{eq:bad_control_proof1}
\mathrm{Re}\left(\frac{\rme^{-\rmi c}}{(K+\lambda)} \sum_{m=0}^K
  \lambda_m \exp\left( \rmi k \, (\theta_m-\alpha_m) \right)\right) \\
\geq 1- \eta\eqsp.
\end{multline}
%where $\mathrm{Re}(z)$ denotes the real part of the complex number $z$.
Hence:
\begin{align*}
1-\eta &\leq \frac{1}{K+\lambda}\sum_{m=0}^K \lambda_m \cos\bigl(k(\theta_m-\alpha_m-c)\bigr)
\\
&\leq \frac{1}{K+\lambda}\sum_{m=0}^K \lambda_m \bigl(1-\gamma_0 k^2(\theta_m-\alpha_m-c)^2\bigr),
\end{align*}
and \eqref{sqErr} follows. Denote by $N$ the number of curves in the block whose alignment error is
``far'' from $c$ (up to a $2\pi$ factor):
\begin{equation*}
N \eqdef \sum_{m=0}^K\boldsymbol1 \left\{ |\theta_m-\alpha_m - c| \geq
  \eta^\delta \right\} \eqsp,
\end{equation*}
and assume, for simplicity, that the $N$ last curves are the misaligned curves.
Equation~(\ref{eq:bad_control_proof1}) implies
\begin{align}\allowdisplaybreaks
 1-\eta
%\nonumber\\
& \leq \frac{1}{K+\lambda}
\sum_{m=0}^{K-N-1}\lambda_m \cos(k(\theta_m-\alpha_m - c))
\nonumber\\
& ~~~+
  \frac{1}{K+\lambda}\sum_{m=K-N}^K \lambda_m \cos(k(\theta_m-\alpha_m - c))
  \nonumber\\
& \leq \frac{K+\lambda-N}{K+\lambda}
\nonumber%\\& ~~~
+ \frac{N}{K} (1-\gamma_0 k^{2\delta}
\eta^{2\delta})
\\
& = 1- \frac{N}{K+\lambda}\gamma_0 k^{2\delta} \eta^{2\delta} \eqsp .\label{eq:bad_control_proof2}
\end{align}
Equation~(\ref{eq:bad_control_proof2}) leads to
$$
N \leq \frac{K+\lambda}{\gamma_0 k^{2\delta}}\eta^{1-2\delta} \eqsp ,
$$
which completes the proof.

\subsection{Proof of Proposition~\ref{prop:choice_lambda}}
\label{sec:proof_choice_lambda}

Assume that $|c| > \eta^{\delta}$; since
$\lambda$ is assumed to be an integer, we can see this weighting
parameter as the artificial addition of $\lambda-1$ reference
curves. Since $\alpha_0 = \theta_0 \eqdef 0$, in that case, $|\theta_0
- \alpha_0 -c| > \eta^\delta$, thus giving
$$
\frac{N}{K+\lambda} > \frac{\lambda}{K + \lambda} \geq
\gamma \eta^{1-2\delta} \eqsp,
$$
which would contradict
Proposition~\ref{prop:bad_control_deterministic}. Therefore, we get
that
$|c| \leq \eta^\delta$.
%See Section~\ref{sec:choice_lambda}
 % version 2.0

%\bibliographystyle{gNST}
\bibliographystyle{IEEEtran}
\bibliography{MARSUabrv,IEEEabrv,Bibliographie}
\end{document}